\documentclass[]{llncs}
	
\usepackage{amssymb}
\usepackage{graphicx}

\usepackage{url}

\newcommand{\TL}{\mathcal{T_L}\xspace}

\usepackage{graphicx} 
\usepackage{subfigure}

\usepackage{stmaryrd}
\usepackage{amsmath, amssymb, amsfonts}
\usepackage{bm}
\usepackage{gdefs}
\usepackage{mathpartir}
\usepackage{array}

\usepackage{tikz}
\usepackage{tikz-qtree}
\usetikzlibrary{arrows,positioning,shapes,patterns}
\tikzset{
    >=stealth',
    boxx/.style={
           rectangle,
           draw=black,
           text width=6.5em,
           minimum height=2em,
           text centered},
}
\usetikzlibrary{calc}
\usepackage{forest}

\usepackage{mathtools}

\DeclarePairedDelimiter\abs{\lvert}{\rvert}%
\DeclarePairedDelimiter\norm{\lVert}{\rVert}%

\makeatletter
\let\oldabs\abs
\def\abs{\@ifstar{\oldabs}{\oldabs*}}
\let\oldnorm\norm
\def\norm{\@ifstar{\oldnorm}{\oldnorm*}}
\makeatother

\usepackage{booktabs}
\newcommand{\ra}[1]{\renewcommand{\arraystretch}{#1}}

\usepackage{listings}

\lstdefinelanguage{JavaScript}{
  keywords={typeof, try, new, true, false, catch, function, return, null, catch, switch, var, if, in, while, do, else, case, break},
  keywordstyle=\color{blue}\bfseries,
  ndkeywords={class, export, boolean, throw, implements, import, this},
  ndkeywordstyle=\color{darkgray}\bfseries,
  identifierstyle=\color{black},
  sensitive=false,
  comment=[l]{//},
  morecomment=[s]{/*}{*/},
  commentstyle=\color{purple}\ttfamily,
  stringstyle=\color{cyan}\ttfamily,
  morestring=[b]',
  morestring=[b]",
  moredelim=[is][\underbar]{_}{_}
}

\usepackage{bold-extra}

\usepackage[noend]{algorithm2e}

\usepackage{wrapfig}

\usepackage{enumitem}

\usepackage{ifthen}
\newcommand{\TODO}[1]{ {\color{blue} \bf #1} }

\newcommand{\ignore}[1]{}

\newcommand{\DONE}[1]{}
\newcommand{\ie}{i.e.}

\newcommand{\eg}{e.g.}
\newcommand{\lemref}[1]{Lemma~\ref{Lm:#1}}
\newcommand{\figref}[1]{Fig.~\ref{Fi:#1}}
\newcommand{\defref}[1]{Definition~\ref{De:#1}}

\newcommand{\tabref}[1]{Table~\ref{Ta:#1}}
\newcommand{\secref}[1]{Section~\ref{Se:#1}}

\newcommand{\appref}[1]{Appendix~\ref{Se:#1}}

\newcommand{\algref}[1]{Algorithm~\ref{Alg:#1}}

\newcommand{\eqtref}[1]{Equation~\ref{Eq:#1}}

\newcommand{\lemlabel}[1]{\label{Lm:#1}}
\newcommand{\figlabel}[1]{\label{Fi:#1}}
\newcommand{\deflabel}[1]{\label{De:#1}}

\newcommand{\tablabel}[1]{\label{Ta:#1}}
\newcommand{\seclabel}[1]{\label{Se:#1}}

\newcommand{\alglabel}[1]{\label{Alg:#1}}
\newcommand{\eqtlabel}[1]{\label{Eq:#1}}

\newcounter{programlinenumber}

\newboolean{TR}
\setboolean{TR}{false}
\ifthenelse{\boolean{TR}}{

\newcommand{\TrOnly}[1]{#1}
\newcommand{\SubOnly}[1]{}
\newcommand{\TrOnlyInFootnote}[1]{#1}
\newcommand{\TrOnlyInTable}[1]{#1}}
{

\newcommand{\TrOnly}[1]{}
\newcommand{\SubOnly}[1]{#1}
\newcommand{\TrOnlyInFootnote}[1]{}
\newcommand{\TrOnlyInTable}[1]{}}

\DeclareMathOperator*{\argmin}{arg\,min}
\DeclareMathOperator*{\argmax}{arg\,max}

\newcommand{\nats}{\mathbb{N}}

\newcommand{\true}{\emph{true}}
\newcommand{\false}{\emph{false}}

\newcommand{\hiddentext}[1]{}

\newcommand{\scode}[1]{\texttt{#1}}
\newcommand{\scodevar}[1]{{\color{blue!60!black}\texttt{#1}}}

\newcommand{\powerset}{\raisebox{.15\baselineskip}{\Large\ensuremath{\wp}}}

\definecolor{graybck}{gray}{0.9}

\newcommand{\mlvec}[1]{\bm{#1}}

\newcommand{\dataset}{\mathcal{D}}

\newcommand{\programs}{P}

\newcommand{\lang}{\ensuremath{\mathcal{L}}}

\newcommand{\analysis}{pa}

\newcommand{\tlang}{\ensuremath{\mathcal{L}_{pt}}}
\newcommand{\alang}{\ensuremath{\mathcal{L}_{alloc}}}

\newcommand{\node}{n}
\newcommand{\tree}{t}
\newcommand{\statements}{s}

\newcommand{\bcode}[1]{\texttt{\textbf{#1}}}
\newcommand{\red}[1]{{\color{red!50!black}{#1}}}
\newcommand{\blue}[1]{{\color{blue!50!black}{#1}}}

\newcommand{\ConcSem}[1]{\llbracket #1 \rrbracket}

\begin{document}

\mainmatter  

\title{Learning a Static Analyzer from Data}

\titlerunning{Learning a Static Analyzer from Data}

\author{Pavol Bielik \and Veselin Raychev \and Martin Vechev}

\authorrunning{Pavol Bielik \and Veselin Raychev \and Martin Vechev}

\institute{
Department of Computer Science, ETH Z{\"u}rich, Switzerland\\
\{pavol.bielik, veselin.raychev, martin.vechev\}@inf.ethz.ch
}

\maketitle

\vspace{-0.7em}
\begin{abstract}
To be practically useful, modern static analyzers must precisely model the effect of both, statements in the programming language as well as frameworks used by the program under analysis. While important, manually addressing these challenges is difficult for at least two reasons: (i) the effects on the overall analysis can be non-trivial, and (ii) as the size and complexity of modern libraries increase, so is the number of cases the analysis must handle.

In this paper we present a new, automated approach for creating static analyzers: instead of manually providing the various inference rules of the analyzer, the key idea is to learn these rules from a dataset of programs. Our method consists of two ingredients: (i) a synthesis algorithm capable of learning a candidate analyzer from a given dataset, and (ii) a counter-example guided learning procedure which generates new programs beyond those in the initial dataset, critical for discovering corner cases and ensuring the learned analysis generalizes to unseen programs.

We implemented and instantiated our approach to the task of learning JavaScript static analysis rules for a subset of points-to analysis and for allocation sites analysis. These are challenging yet important problems that have received significant research attention. We show that our approach is effective: our system automatically discovered practical and useful inference rules for many cases that are tricky to manually identify and are missed by state-of-the-art, manually tuned analyzers.
\end{abstract}

\section{Introduction}\seclabel{intro}
Static analysis is a fundamental method for automating program reasoning with a myriad of applications in verification, optimization and bug finding. While the theory of static analysis is well understood, building an analyzer for a practical language is a highly non-trivial task, even for experts. This is because one has to address several conflicting goals, including: (i) the analysis must be scalable enough to handle realistic programs, (ii) be precise enough to not report too many false positives, (iii) handle tricky corner cases and specifics of the particular language (e.g., JavaScript), (iv) decide how to precisely model the effect of the environment (e.g., built-in and third party functions), and other concerns. Addressing all of these manually, by-hand, is difficult and can easily result in suboptimal static analyzers, hindering their adoption in practice.

\paragraph{Problem statement}
The goal of this work is to help experts design robust static analyzers, faster, by automatically learning key parts of the analyzer from data.

We state our learning problem as follows: given a domain-specific language $\mathcal{L}$ for describing analysis rules (i.e., transfer functions, abstract transformers), a~dataset $\dataset$ of programs in some programming language (e.g., JavaScript), and an abstraction function $\alpha$ that defines how concrete behaviors are abstracted, the goal is to learn an analyzer $\analysis \in \mathcal{L}$ (i.e., the analysis rules) such that programs in $\dataset$ are analyzed as precisely as possible, subject to $\alpha$.

\paragraph{Key challenges}
There are two main challenges we address in learning static analyzers. First, static analyzers are typically described via rules (i.e., type inference rules, abstract transformers), designed by experts, while existing general machine learning techniques such as support vector machines and neural networks only produce weights over feature functions as output. If these existing techniques were applied to program analysis~\cite{JSNICE,MayurAnalysis}, the result would simply be a~(linear) combination of existing rules and no new interesting rules would be discovered. Instead, we introduce domain-specific languages for describing the analysis rules, and then learn such analysis rules (which determine the analyzer) over these languages.

The second and more challenging problem we address is how to avoid learning a static analyzer that works well on some training data $\dataset$, but fails to generalize well to programs outside of $\dataset$ -- a problem known in machine learning as \emph{overfitting}. We show that standard techniques from statistical learning theory~\cite{StatisticalLearningTheory} such as \emph{regularization} are insufficient for our purposes.
The idea of regularization is that picking  a simpler model minimizes the expected error rate on unseen data, but a simpler model also contradicts an important desired property of static analyzers to \emph{correctly handle tricky corner cases}. We address this challenge via a~counter-example guided learning procedure that leverages program semantics to generate new data (i.e., programs) for which the learned analysis produces wrong results and which are then used to further refine it.
To the best of our knowledge, we are the first to replace model regularization with a~counter-example guided procedure in a machine learning setting with large and noisy training datasets.

We implemented our method and instantiated it for the task of learning production rules of realistic analyses for JavaScript. We show that the learned rules for points-to and for allocation site analysis are indeed interesting and are missed by existing state-of-the-art, hand crafted analyzers (e.g., Facebook's Flow~\cite{facebookflow}) and TAJS (e.g., \cite{TAJS}).

Our main contributions are:

\begin{itemize}
\item A method for learning static analysis rules from a dataset of programs. To ensure that the analysis \emph{generalizes} beyond the training data we carefully generate counter-examples to the currently learned analyzer using an oracle.

\item A decision-tree-based algorithm for learning analysis rules from data that \emph{learns to overapproximate} when the dataset cannot be handled precisely.

\item An end-to-end implementation of our approach and an evaluation on the challenging problem of learning tricky JavaScript analysis rules. We show that our method produces interesting analyzers which generalize well to new data (i.e. are sound and precise) and handle many tricky corner cases.

\end{itemize}

\tikzset{
    >=stealth',
    punkt/.style={
           rectangle,
           rounded corners,
           draw=black, very thick,
           text width=6.5em,
           minimum height=2em,
           text centered},
    pil/.style={
           ->,
           very thick,
           shorten <=2pt,
           shorten >=2pt,}
}

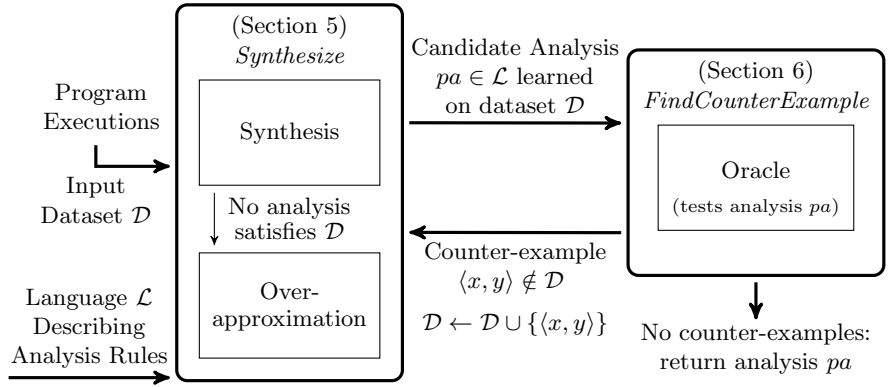
\begin{figure*}[t]
\centering
\begin{tikzpicture}

 \node[punkt, minimum height=5cm, minimum width=3cm] (alg) at (8, 8) {};
 \node[punkt, minimum height=3cm, minimum width=3.4cm] (oracle) at (14.2, 8.4) {};

\begin{scope}[xshift=0.4cm,yshift=1.4cm]
 \node at (13.8, 8.2) {(\secref{oracle})};
 \node at (13.8, 7.8) {$\textit{FindCounterExample}$};
\end{scope}

 \node at (8, 10.2) {(\secref{learning})};
 \node at (8, 9.8) {$\textit{Synthesize}$};

\begin{scope}[xshift=6.2cm,yshift=-0.6cm]
 \draw (6.7, 9.5) rectangle (9.3, 8.1);
  \node at (8, 8.9) {Oracle};
  \node at (8, 8.4) {\scriptsize (tests analysis $\analysis$)};
\end{scope}

 \draw (6.8, 9.5) rectangle (9.2, 8.1);
 \node at (8, 8.8) {Synthesis};

 \draw (6.8, 7.2) rectangle (9.2, 5.8);
 \node at (8, 6.7) {Over-};
 \node at (8, 6.3) {approximation};

 \draw [->] (7.0, 8.0) -- (7.0, 7.3);
 \node at (8, 7.8) {No analysis};
 \node at (8, 7.5) {satisfies $\dataset$};

\draw[pil] (oracle.south) -- ([yshift=-18pt]oracle.south);
\node[below=0.5cm of oracle, align=center] {No counter-examples:\\ return analysis $\analysis$};

\draw[pil] ([yshift=26.5pt]alg.east) -- node[above, align=center] {Candidate Analysis \\ $\analysis \in \mathcal{L}$ learned \\ on dataset $\dataset$} ([yshift=15pt]oracle.west);
\draw[pil] ([yshift=-26.5pt]oracle.west) -- node[below, align=center] {Counter-example \\ $\langle x, y\rangle \notin \dataset$ \vspace{0.1cm} \\ $\dataset \leftarrow \dataset \cup \{\langle x, y\rangle \}$ } ([yshift=-15pt]alg.east);

\draw[pil] ([xshift=-65pt, yshift=-70pt]alg.west) -- node[above, align=center] {\\ Language $\mathcal{L}$\\ Describing\\ Analysis Rules} ([yshift=-70pt]alg.west);

\draw[pil] ([xshift=-30pt, yshift=20pt]alg.west) |- node[above, yshift=-25pt, align=center] {Input\\ Dataset $\dataset$} ([yshift=10pt]alg.west);

\node[align=center] at (5.5, 9.15) {Program \\ Executions};

\end{tikzpicture}
\caption{Overview of our approach to learning static analysis rules from data consisting of three components -- a language~$\mathcal{L}$ for describing the rules, a learning algorithm and an oracle -- that interact in a counter-example based refinement loop.}
\figlabel{approach}
\end{figure*}
\section{Our Approach}

We begin by describing components of our learning approach as shown in~\figref{approach}.

\paragraph{Obtaining training data $\dataset$}
Our learning approach uses dataset of examples $\dataset = \lbrace \langle x^j, y^j \rangle \rbrace_{j=1}^N$ consisting of pairs $ \langle x^j, y^j \rangle $ where $x^j$ is a program and $y^j$ is the desired output of the analysis when applied to $x^j$.
In general, obtaining such labeled training data for machine learning purposes is a tedious task. In our setting, however, this process can be automated because: (i) in static analysis, there is a well understood notion of correctness, namely, the analyzer must \emph{approximate} (in the sense of lattice ordering) the concrete program behaviors, and (ii) thus, we can simply run a large amount of programs in a given programming language with some inputs, and obtain a subset of the concrete semantics for each program.
We note that our learning method is independent of how the labels are obtained. For example, the labels $y^j$ can be obtained by running static or dynamic analyzers on the programs $x^j$ in $\dataset$ or they can be provided manually.


\paragraph{Synthesizer and Language \lang{}}
To express interesting rules of a static analyzer, we use a loop-free domain-specific language \lang{} with branches (detailed description is provided in \appref{DSLlanguage} and \appref{allocsite}).
The synthesizer then takes as input the dataset $\dataset$ with a~language \lang{} and produces a candidate program analysis $\analysis \in \lang{}$ which correctly handles the pairs in $\dataset$. The synthesizer we propose phrases the problem of learning a static analysis over \lang{} as a problem in learning decision trees over \lang{}. These components are described in \secref{learning}.

\paragraph{Oracle}
Our goal is to discover a program analysis that not only behaves as described by the pairs in the dataset~$\dataset$, but one that generalizes to programs beyond those in $\dataset$. To address this challenge, we introduce the oracle component ($\textit{FindCounterExample}$) and connect it with the synthesizer. This component takes as input the learned analysis $\analysis$ and tries to find another program $x$ for which $\analysis$ fails to produce the desired result $y$. This counter-example $\langle x, y \rangle$ is then fed back to the synthesizer which uses it to generate a new candidate analyzer as illustrated in \figref{approach}. To produce a counter-example, the oracle must have a~way to quickly and effectively test a (candidate) static analyzer. In \secref{oracle}, we present two techniques that make the testing process more effective by leveraging the current set $\dataset$ as well as current candidate analysis $\analysis$ (these techniques for testing a~static analyzer are of interest beyond learning considered in our work).

\paragraph{Counter-example guided learning}
To learn a static analyzer $\analysis$, the synthesizer and the oracle are linked together in a counter-example guided loop. This type of iterative search is frequently used in program synthesis \cite{CombinatorialSketching}, though its instantiation heavily depends on the particular application task at hand. In our setting, the examples in $\dataset$ are programs (and not say program states) and we also deal with notions of (analysis) approximation. This also means that we cannot directly leverage off-the-shelf components (e.g., SMT solvers) or existing synthesis approaches. Importantly, the counter-example guided approach employed here is of interest to machine learning as it addresses the problem of overfitting with techniques beyond those typically used (e.g., regularization~\cite{StatisticalLearningTheory}, which is insufficient here as it does not consider samples not in the training dataset).

\paragraph{Practical applicability}
We implemented our approach and instantiated it to the task of learning rules for points-to and allocation site analysis for JavaScript code. This is a practical and relevant problem because of the tricky language semantics and wide use of libraries. Interestingly, our system learned inference rules missed by manually crafted state-of-the-art tools, e.g., Facebook's Flow~\cite{facebookflow}. 

\section{Overview}\seclabel{overview}
This section provides an intuitive explanation of our approach on a simple points-to analysis for JavaScript. Assume we are learning the analysis from one training data sample given in \figref{paexample} (a). It consists of variables \scodevar{a}, \scodevar{b} and \scodevar{b} is assigned an object $s_0$. Our goal is to learn that \scodevar{a} may also point to that same object~$s_0$.

\begin{figure}
\centering

\begin{tikzpicture}

\begin{scope}[xshift=-2cm,yshift=0cm]
\node[right] at (1.0, 4.0) {\scode{var \scodevar{b} = \{\};} // empty object $s_0$ };
\node[right] at (1.0, 3.7) {\scode{\scodevar{a} = \scodevar{b};}};

\node[right] at (1.0, 3.0) {Expected points-to set};
\node[right] at (1.0, 2.6) {$\dataset$ = \{($\scodevar{a} \rightarrow \{s_0\}$)\}};

\node at (3.0, 1.95) {(a) Training data};
\end{scope}

\begin{scope}[xshift=3.3cm,yshift=2.7cm]

\node at (4.0, 0.8) {
\begin{minipage}{.4\textwidth}
\begin{forest}
  for tree={
    font=\ttfamily,
    grow'=0,
    child anchor=west,
    parent anchor=south,
    anchor=west,
    calign=first,
    edge path={
      \noexpand\path [draw, \forestoption{edge}]     (!u.south west) +(10pt,2pt) |- node[fill,inner sep=1.25pt] {} (.child anchor)\forestoption{edge label};
    },
    before typesetting nodes={
      if n=1
        {insert before={[,phantom]}}
        {}
    },
    fit=band,
    s sep=2pt,
    before computing xy={l=15pt},
  }
[
    [VarDeclaration:\scodevar{b}
      [ObjectExpression:\scode{\{\}}]
    ]
    [Assignment
      [Identifier:\scodevar{a}]
      [Identifier:\scodevar{b}]
    ]
]
\end{forest}
\end{minipage}
};

\node at (3.9, -0.75) {(b) Abstract syntax tree (AST) representation of (a)};

\end{scope}

\begin{scope}[xshift=1.8cm,yshift=0.8cm]
\node at (0, 0) {
\begin{minipage}{.4\textwidth}

\begin{align*}
&\quad f_{desired}(x) ::=  \\
&y\quad \textbf{if~there~is~}\texttt{Assignment}(x, y) \\
&y\quad \textbf{if~there is~}\texttt{VarDeclaration:\scodevar{x}}(y) \\
&\bottom\quad \textbf{otherwise}
\end{align*}

\end{minipage}
};
\end{scope}

\begin{scope}[xshift=7.9cm,yshift=0.8cm]

\node at (0, 0) {
\begin{minipage}{.4\textwidth}

\begin{align*}
&\quad f_{overfit}(x) ::=  \\
&y\quad \textbf{if~} y \textbf{~is~}\texttt{VarDeclaration:\scodevar{y}}\textbf{~preceding~}x \\
&y\quad \textbf{if~there is~}\texttt{VarDeclaration:\scodevar{x}}(y) \\
&\bottom\quad \textbf{otherwise}
\end{align*}

\end{minipage}
};

\node at (-3.0, -1.5) {(c) Learned functions to resolve points-to queries from (a)};

\end{scope}

\end{tikzpicture}
\caption{Example data for learning points-to analysis.}
\figlabel{paexample}
\end{figure}
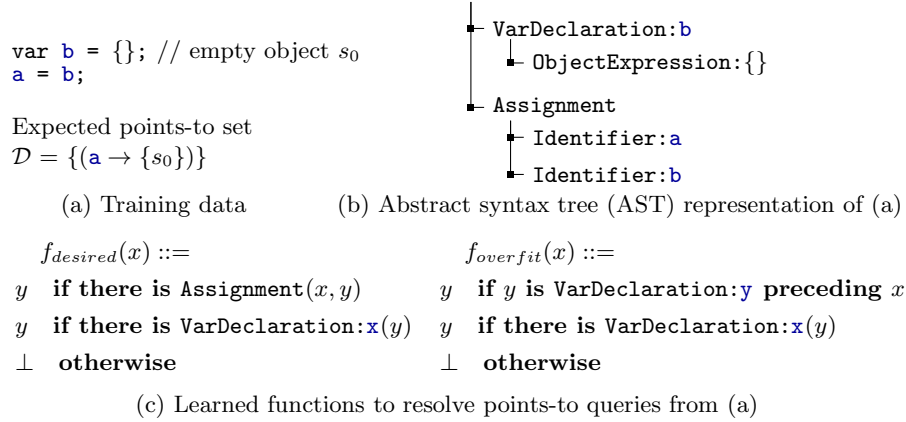
Points-to analysis is typically done by applying inference rules until fixpoint. An example of an inference rule modeling the effect of assignment is:

\[
\footnotesize
\inferrule*[Right=\footnotesize \textsc{$~$[Assign]}]
{\scode{VarPointsTo(}v_2, h\scode{)} \+ \scode{Assignment(}v_1, v_2\scode{)}}
{\scode{VarPointsTo(}v_1, h\scode{)}}
\]

\noindent
This rule essentially says that if variable $v_2$ is assigned to $v_1$ and $v_2$ may point to an object $h$, then the variable $v_1$ may also point to this object $h$.

\paragraph{Domain specific language (DSL) for analysis rules:} Consider the following general shape of inference rules:
\[
\footnotesize
\inferrule*[Right=\footnotesize \textsc{$~$[General]}]
{\scode{VarPointsTo(}v_2, h\scode{)} \+ v_2 = f(v_1)}
{\scode{VarPointsTo(}v_1, h\scode{)}}
\]

Here, the function $f$ takes a program element (a variable) and returns another program element or $\bottom$. The rule says: use the function $f$ to find a variable $v_2$ whose points-to set will be used to determine what $v_1$ points to. The \textsc{Assign} rule is an instance of the \textsc{General} rule that can be implemented by traversing the AST and checking if the parent node of $x$ is of type \texttt{Assignment} and if $x$ is its first child. In this case, the right sibling of $x$ is returned. Otherwise $f$ returns~$\bottom$.

\paragraph{Problem statement}
The problem of learning a points-to analysis can now be stated as follows: find an analysis $\analysis \in  \lang{}$ such that when analyzing the programs in the training data $\dataset$, the resulting points-to set is as outlined in $\dataset$.

\paragraph{The overfitting problem}
Consider \figref{paexample} (b) which shows the AST of our example.
In addition to \textsc{Assign}, we need to handle the case of variable initialization (first line in the program).
Note that the dataset $\dataset$ does not uniquely determine the best function $f$. In fact, instead of the desired one $f_{desired}$, other functions can be returned such as $f_{overfit}$ shown in \figref{paexample} (c).
This function inspects the statement prior to an assignment instead of at the assignment itself and yet it succeeds to produce the correct analysis result on our dataset $\dataset$. However, this is due to the specific syntactic arrangement of statements in the training data $\dataset$ and may not generalize to other programs, beyond those in $\dataset$.

\paragraph{Our solution}
To address the problem of overfitting to $\dataset$, we propose a counter-example guided procedure that biases the learning towards semantically meaningful analyses. That is, the oracle tests the current analyzer and tries to find a~counter-example on which the analysis fails. Our strategy to generating candidate programs is to modify the programs in $\dataset$ in ways that can change both the syntax and the semantics of those programs. As a result, any analysis that depends on such properties would be penalized in the next iteration of $\textit{Synthesize}$. As we show in the evaluation, our approach results in a much faster oracle than if we had generated programs blindly. This is critical as faster ways of finding counter-examples increase the size of the search space we can explore, enabling us to discover interesting analyzers in reasonable time.

For example, a possible way to exclude $f_{overfit}$ is to insert an unnecessary statement (e.g., $\scode{var \scodevar{c} = 1}$) before the assignment $\scodevar{a} = \scodevar{b}$ in \figref{paexample} (a). Here, the analysis defined by $f_{overfit}$ produces an incorrect points-to set for variable \scodevar{a} (as it points-to the value~$1$ of variable \scodevar{c}). Once this sample is added to $\dataset$, $f_{overfit}$ is penalized as it produces incorrect results and the next iteration will produce a different analysis until eventually the desired analysis $f_{desired}$ is returned.

\paragraph{Correctness of the approach}
Our method produces an analyzer that is guaranteed to be sound w.r.t to all of the examples in $\dataset$. Even if the analyzer cannot exactly satisfy all examples in $\dataset$, the synthesis procedure always returns an \emph{over-approximation} of the desired outputs. That is, when it cannot match the target output exactly, $\textit{Synthesize}$ learns to approximate (e.g., can return $\top$ in some cases). A formal argument together with a discussion on these points is provided in \secref{learning}. However, our method is not guaranteed to be sound for all programs in the programming language. We see the problem of certifying the analyzer as orthogonal and complementary to our work: our method can be used to predict an analyzer which is likely correct, generalize well, and to sift through millions of possibilities quickly, while a follow-up effort can examine this analyzer and decide whether to accept it or even fully verify it. Here, an advantage of our method is that the learned analyzer is expressed as a program, which can be easily examined by an expert (we show examples of learned analyzers in \appref{examples}), as opposed to standard machine learning models where interpreting the result is nearly impossible and therefore difficult to verify with standard methods.

\section{Checking Analyzer Correctness}\seclabel{analysisCorrectness}
In this section, following \cite{Cousot:1977}, we briefly discuss what it means for a (learned) analyzer to be correct.
The concrete semantics of a program $p$ include all of $p$'s concrete behaviors and are captured by a function $\ConcSem{p} \colon \mathbb{N} \rightarrow \powerset{(\mathcal{C})}$. This function associates a set of possible concrete states in $\mathcal{C}$ with each position in the program $p$, where a position can be a program counter or a node in the program's AST.

A static analysis $\analysis$ of a program $p$ computes an abstract representation of the program's concrete behaviors, captured by a function $\analysis(p) \colon \mathbb{N} \rightarrow \mathcal{A}$ where $(\mathcal{A}, \sqsubseteq)$ is typically an abstract domain, usually a lattice of abstract facts equipped with an ordering $\sqsubseteq$ between facts. An abstraction function $\alpha \colon \powerset{(\mathcal{C})} \rightarrow \mathcal{A}$ then establishes a connection between the concrete behaviors and the abstract facts. It defines how a set of concrete states in $\mathcal{C}$ is abstracted into an abstract element in $\mathcal{A}$. The function is naturally lifted to work point-wise on a set of positions in~$\mathbb{N}$ (used in the definition below).

\begin{definition}[Analysis Correctness]\deflabel{correctness}
A static analysis $\analysis$ is correct if:
\begin{equation}
\forall p \in \TL. ~\alpha(\ConcSem{p}) \sqsubseteq \analysis(p)
\end{equation}
\end{definition}

Here $\TL$ denotes the set of all possible programs in the target programming language ($\TL$). That is, a static analysis is correct if it over-approximates the concrete behaviors of the program according to the particular lattice ordering.

\subsection{Checking Correctness}\seclabel{PAtesting}
One approach for checking the correctness of an analyzer is to try and automatically verify the analyzer itself, that is, to prove the analyzer satisfies \defref{correctness} via sophisticated reasoning (e.g., as the one found in \cite{Giacobazzi:2015}).
Unfortunately, such automated verifiers do not currently exist (though, coming up with one is an interesting research challenge) and even if they did exist, it is prohibitively expensive to place such a~verifier in the middle of a~counter-example learning loop where one has to discard thousands of candidate analyzers quickly. Thus, the correctness definition that we use in our approach is as follows:

\begin{definition}[Analysis Correctness on a Dataset and Test Inputs]\deflabel{testedcorrectness}
A static analysis $\analysis$ is correct w.r.t to a dataset of programs $\programs$ and test inputs $\textit{ti}$~if:
\begin{equation}
\forall p \in \programs. ~\alpha(\ConcSem{p}_{\textit{ti}}) \sqsubseteq \analysis(p)
\end{equation}
\end{definition}

The restrictions over \defref{correctness} are: the use of a set $\programs \subseteq \TL$ instead of $\TL$ and $\ConcSem{p}_{ti}$ instead of $\ConcSem{p}$. Here, $\ConcSem{p}_{\textit{ti}} \subseteq \ConcSem{p}$ denotes a subset of a program $p$'s behaviors obtained after running the program on some set of test inputs $\textit{ti}$.

The advantage of this definition is that we can automate its checking. We run the program $p$ on its test inputs $\textit{ti}$ to obtain $\ConcSem{p}_{\textit{ti}}$ (a finite set of executions) and then apply the function $\alpha$~on the resulting set. To obtain $\analysis(p)$, we run the analyzer $\analysis$ on $p$; finally, we compare the two results via the inclusion operator~$\sqsubseteq$. 
\section{Learning Analysis Rules}\seclabel{learning}
We now present our approach for learning static analysis rules from examples.

\subsection{Preliminaries}
Let $\dataset = \lbrace \langle x^j, y^j \rangle \rbrace_{j=1}^N$ be a dataset of programs from a target language $\TL$ together with outputs that a program analysis should satisfy. That is, $x^j \in \TL$ and $y^j$ are the outputs to be satisfied by the learned program analysis.

\begin{definition}[Analysis Correctness on Examples]\deflabel{correctnessExamples}
We say that a static analysis $\analysis \in \lang$ is correct on $\dataset = \lbrace \langle x^j, y^j \rangle \rbrace_{j=1}^N$ if:
\begin{equation}
\forall j \in 1 \dots N~~.~ ~y^j \sqsubseteq \analysis(x^j)
\end{equation}
\end{definition}
This definition is based on \defref{testedcorrectness}, except that the result of the analysis is provided in $\dataset$ and need not be computed by running programs on test inputs.

Note that the definition above does not mention the precision of the analysis $\analysis$ but is only concerned with soundness.
To search for an analysis that is both sound and precise and avoids obvious, but useless solutions (e.g., always return $\top$ element of the lattice $(\mathcal{A}, \sqsubseteq)$), we define a precision metric.

\paragraph{Precision metric}
First, we define a function $r \colon \TL \times \mathcal{A} \times{} \lang \rightarrow \mathbb{R}$ that takes a program in the target language, its desired program analysis output and a~program analysis and indicates if the result of the analysis is exactly as desired:
\begin{equation}
r(x, y, \analysis) ~~=~~ \textbf{if~} (y \ne \analysis(x)) \textbf{~then~} 1 \textbf{~else~} 0
\end{equation}

\noindent
We define a function $\textit{cost}$ to compute precision on the full dataset $\dataset$ as follows:
\begin{equation}\eqtlabel{cost}
\textit{cost}(\dataset, \analysis) = \sum_{\langle x,y \rangle \in \dataset} r(x, y, pa)
\end{equation}

\noindent
Using the precision metric in \eqtref{cost}, we can state the following lemma:
\begin{lemma}\lemlabel{zerocost}
For a program analysis $\analysis \in \lang$ and a dataset $\dataset$,
if $\textit{cost}(\dataset, \analysis) = 0$, then the analysis is correct according to \defref{correctnessExamples}.
\end{lemma}
\noindent \textbf{Proof:} The proof is direct. Because $\textit{cost}(\dataset, \analysis) = 0$ and $r$ is positive, then for every  $\langle x, y \rangle \in \dataset$,
$r(x, y, \analysis) = 0$. This means that $y=\analysis(x)$ and so $y \sqsubseteq \analysis(x)$, which is as defined in \defref{correctnessExamples}.
\qed

\subsection{Problem Formulation}\seclabel{LearnProblem}
Given a language $\lang$ that describes analysis inference rules (i.e., abstract transformers) and a dataset $\dataset$ of programs with the desired analysis results, the $\textit{Synthesize}$ procedure should return a program analysis $\analysis \in \lang$ such that:
\begin{enumerate}
  \item $\analysis$ is correct on the examples in $\dataset$ (\defref{correctnessExamples}), and
  \item $\textit{cost}(\dataset, pa)$ is minimized.
\end{enumerate}

The above statement essentially says that we would like to obtain a sound analysis which also minimizes the over-approximation that it makes. As the space of possible analyzers can be prohibitively large, we discuss a restriction on the language $\lang$ and give a procedure that efficiently searches for an analyzer such that correctness is enforced and $\textit{cost}$ is (approximately) minimized.

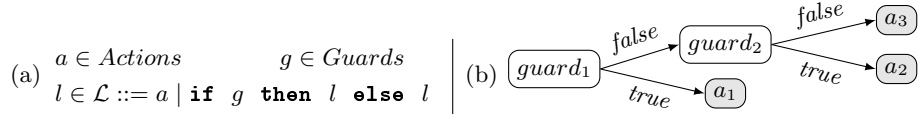
\begin{figure}[t]
\centering
\tikzset{
  treenode/.style = {shape=rectangle, rounded corners,
                     draw, align=center,
                     fill=gray!20},
  internal/.style     = {treenode, font=\small, fill=white!30},
  env/.style      = {treenode, font=\small},
}

\begin{tikzpicture}
  [
    grow                    = right,
    sibling distance        = 2em,
    level distance          = 7em,
    edge from parent/.style = {draw, -latex},
    every node/.style       = {font=\footnotesize},
    sloped
  ]

\node[right] at (-4.1, 4.5) {(a)};
\node[right] at (-3.5, 4.7) {
\begin{minipage}{.35\textwidth}
\begin{gather*}
a \in Actions \quad \quad \quad \quad g \in Guards \\
l \in \lang :: = a \mid \bcode{if}~~g~~\bcode{then}~~l ~~\bcode{else}~~l
\end{gather*}
\end{minipage}
};

\draw (1.9, 5.0) -- (1.9, 4.0);

\node[right] at (1.95, 4.5) {(b)};

  \node [internal] at (3.25, 4.6) {$guard_1$}
    child { node [env] {$a_1$}
      edge from parent node [below] {$\true$}
    }
    child { node [internal] {$guard_2$}
      child { node [env] {$a_2$}
        edge from parent node [below] {$\true$} }
      child { node [env] {$a_3$}
        edge from parent node [above] {$\false$} }
      edge from parent node [above] {$\false$}
    };
\end{tikzpicture}
\vspace{-1.7em}
\caption{(a) Syntax of a template language $\lang$ with branches for expressing analysis rules. (b) Example of a function from the $\lang$ language shown as a decision tree.
}
\figlabel{template}
\end{figure}
\subsection{Language Template for Describing Analysis Rules}\seclabel{lang}
A template of the language $\lang$ for describing analysis rules is shown in \figref{template} (a). The template is simple and contains actions and guards that are to be instantiated later. The statements in the language are either an action or a~conditional \bcode{if-then-else} statements that can be applied recursively.

An analysis rule of a static analyzer are expressed as a function built from statements in $\lang$. As usual, the function is executed until a~fixed point~\cite{Cousot:1977}. The semantics of the \bcode{if} statements in $\analysis$ is standard: guards are predicates (side-effect free) that inspect the program being analyzed and depending on their truth value, the corresponding branch of the \bcode{if} statement is taken. The reason such \bcode{if} statements are interesting is because they can express analysis rules such as the ones of our running example in \figref{paexample}. 

We provide a formal semantics and detailed description of how the language~$\lang$ is instantiated for learning points-to and allocation site analysis in \appref{DSLlanguage} and \appref{allocsite} respectively.

\subsection{ID3 Learning for a Program Analyzer}\seclabel{id3}
A key challenge in learning program analyzers is that the search space of possible programs over $\lang$ is massive as the number of possible combinations of branches and subprograms is too large. However, we note that elements of $\lang$ can be represented as trees where internal nodes are guards of \bcode{if} statements and the leafs are actions as shown in \figref{template} (b). Using this observation we can phrase the problem of learning an analyzer in $\lang$ as the problem of learning a decision tree, allowing us to adapt existing decision tree algorithms to our setting.

Towards that, we extend the ID3~\cite{ID3} algorithm to handle action programs in the leafs and to enforce correctness of the resulting analysis $\analysis \in \lang$. Similarly to ID3, our algorithm is a greedy procedure that builds the decision tree in a~top-down fashion and locally maximizes a metric called information gain.

Our learning shown in \algref{LearnAlgo} uses three helper functions that we define next.
First, the $\textit{genAction}$ function returns best analysis $a_{best}$ for a dataset $\dataset$:
\begin{equation}
a_{best} = \textit{genAction}(\dataset) = \argmin_{a \in Actions} cost(\dataset, a)
\end{equation}
That is, $\textit{genAction}$ returns the most precise program analysis consisting only of $Actions$ (as we will see later, an action is just a sequence of statements, without branches).
If $a_{best}$ is such that $\textit{cost}(\dataset, a_{best})~=~0$, the analysis is both precise and correct (from \lemref{zerocost}), which satisfies our requirements stated in \secref{LearnProblem} and we simply return it. Otherwise, we continue by generating an \bcode{if} statement.

\paragraph{Generating branches}
The ID3 decision tree learning algorithm generates branches based on an information gain metric.
To define this metric, we first use a standard definition of entropy. Let the vector $\mlvec{w} = \langle w_1,..., w_k \rangle$ consist of elements from a~set $C$. Then the entropy $H$ on $\mlvec{w}$ is:
\begin{equation}
H(\mlvec{w}) = -\sum_{c \in C} \frac{\textit{count}(c, \mlvec{w})}{k} \log_2 \Bigg( \frac{\textit{count}(c, \mlvec{w})}{k} \Bigg)
\end{equation}
where $\textit{count}(c, \mlvec{w}) = |~\{ i \in 1 \dots k \mid w_i = c \}~|$.

For a dataset $d \subseteq \dataset$, let $d = \{ x_i, y_i \}_{i=1}^{|d|}$. Then, we define the following vector:
\begin{equation}
\mlvec{w}_d^{a_{best}} = \langle r(x_i, y_i, a_{best}) \mid i \in 1 \dots |d| \rangle
\end{equation}
That is, for every program in $d$, we record if $a_{best}$ is a precise analysis (via the function $r$ defined previously). Let $g \in Guards$ be a predicate that is to be evaluated on a program $x$. Let $\dataset^{g} = \{ \langle x, y \rangle \in \dataset \mid g(x) \}$ and $\dataset^{\neg g} = \dataset \setminus \dataset^{g}$.

The information gain on a set of examples $\dataset$ for analysis $a_{best}$ and predicate guard $g$ is then defined as:
\begin{equation}
IG^{a_{best}}(\dataset, g) = H( \mlvec{w}_{\dataset}^{a_{best}} ) - \frac{|\dataset^{g}|}{|\dataset|} H( \mlvec{w}_{\dataset^{g}}^{a_{best}} ) - \frac{|\dataset^{\neg g}|}{|\dataset|} H( \mlvec{w}_{\dataset^{\neg g}}^{a_{best}} )
\end{equation}

For a given predicate $g$, what the information gain quantifies is how many bits of information about the analysis correctness will be saved if instead of using the imprecise analysis $a_{best}$ directly, we split the dataset with a predicate~$g$.
Using the information gain metric we define $\textit{genBranch}$ as follows:
\begin{equation}
g_{best} = \textit{genBranch}(a_{best}, \dataset) = \argmax^{~~~~~~~~~~~~~~~~\bottom}_{g \in Guards} IG^{a_{best}}(\dataset, g)
\end{equation}
Here, $\argmax^{\bottom}$ is defined to return $\bottom$ if the maximized information gain is $0$, or otherwise to return the guard $g$ which maximizes the information gain.

Back to \algref{LearnAlgo}, if $\textit{genBranch}$ returns a predicate with positive information gain, we split the dataset with this predicate and call $\textit{Synthesize}$ recursively on the two parts. In the end, we return an \bcode{if} statement on the predicate $g$ and the two recursively synthesized analysis pieces.

\SetKwFunction{FRecurs}{FnRecursive}%
\SetKwProg{Fn}{def}{\string:}{}

\setlength{\textfloatsep}{15pt}
\begin{algorithm}[t]
\alglabel{LearnAlgo}

\SetKwRepeat{Do}{do}{while}%
\SetKwProg{myproc}{def}{}{}
\myproc{$\textit{Synthesize}(\dataset)$}{
\KwIn{Dataset $\dataset = \lbrace \langle x^j, y^j \rangle \rbrace_{j=1}^N$}
\KwOut{Program $\analysis \in \lang$}
\SetAlgoShortEnd

  {$a_{best} \leftarrow \textit{genAction}(\dataset)$\\}
  \lIf{$\textit{cost}(\dataset, a_{best}) = 0$}{\Return{$a_{best}$}}
  {$g_{best} \leftarrow \textit{genBranch}(a_{best}, \dataset)$\\}

  \lIf{$g_{best} = \bottom$}{	
  	\Return{$\textit{approximate}(\dataset)$// $\dataset$ are noisy examples}
  }

  $p_1 \leftarrow \textit{Synthesize}(\{ \langle x, y \rangle \in \dataset \mid g_{best}(x)\})$\\
  $p_2 \leftarrow \textit{Synthesize}(\{ \langle x, y \rangle \in \dataset \mid \neg g_{best}(x)\})$\\
	
  \Return{$\bcode{if}~g_{best}~\bcode{then}~p_1~\bcode{else}~p_2$}

}
\vspace{0.5em}
\caption{Learning algorithm for programs from language $\lang$.}
\end{algorithm}

\paragraph{Approximation}
If the information gain is $0$ (i.e. $g_{best} = \bottom$), we could not find any suitable predicate to split the dataset and the analysis $a_{best}$ has non-zero cost. In this case, we define a function $\textit{approximate}$ that returns an approximate, but correct program analysis -- in our implementation we return analysis that loses precision by simply returning $\top$, which is always a correct analysis.

In practice, this approximation does not return $\top$ for the entire analysis, but only for few of the branches in the decision tree, for which the synthesis procedure fails to produce a good program using both $\textit{genAction}$ and $\textit{getBranch}$.

In terms of guarantees, for \algref{LearnAlgo}, we can state the following lemma.
\begin{lemma}\lemlabel{soundness}
The analysis $\analysis \in \lang$ returned by $\textit{Synthesize}$ is correct according to \defref{correctnessExamples}.
\end{lemma}
\noindent The proof of this lemma simply follows the definition of the algorithm and uses induction for the recursion. For our induction base, we have already shown that in case $\textit{cost}(\dataset, a_{best}) = 0$, the analysis is correct. The analysis is also correct if $\textit{approximate}$ is called. In our induction step we use the fact that analyses $p_1$ and $p_2$ from the recursion are correct and must only show that the composed analysis $\bcode{if}~g_{best}~\bcode{then}~p_1~\bcode{else}~p_2$ is also correct.

\section{The Oracle: Testing an Analyzer}\seclabel{oracle}
A key component of our approach is an oracle that can quickly test whether the current candidate analyzer is correct, and if not, to find a counter-example. The oracle takes as an input a candidate analyzer $\analysis$ and the current dataset $\dataset$ used to learn $\analysis$ and outputs a counter-example program on which $\analysis$ behaves incorrectly. More formally, if $\programs_{\dataset} = \{ x \mid \langle x, y \rangle \in \dataset \}$, our goal is to find a counter-example program $p \in \TL$ such that $p \notin \programs_{\dataset}$ and the correctness condition in \defref{testedcorrectness} is violated for the given analysis $\analysis$ and program~$p$. That is, our oracle must generate new programs beyond those already present in $\programs_{\dataset}$.

\paragraph{Key Challenge}
A key problem the oracle must address is to \emph{quickly} find a~counter-example in the search space of all possible programs.
As we show in \secref{evaluation}, finding such a counter-example by blindly generating new programs does not work as the search space of programs in $\TL$ is massive (or even infinite).

\paragraph{Speeding up the search}
We address this challenge by designing a general purpose oracle that \emph{prioritizes} the search in $\TL$ based on ideas inspired by state-of-the-art testing techniques~\cite{SAGE,EmiTesting}.
In particular, we generate new programs by performing modifications of the programs in $\programs_{\dataset}$. These modification are carefully selected by exploiting the structure of the current analysis $\analysis$ in two ways: (i)~to select a program in $\TL$ and the position in that program to modify, and (ii) to determine what modification to perform at this position.

\subsection{Choosing Modification Positions}\seclabel{guiding}

Given a program $x \in \programs_{\dataset}$ and analysis $\analysis$, we prioritize positions that are \emph{read} while executing the program analysis $\analysis$ and changing them would trigger different \emph{execution path} in the analyzer $\analysis$ itself (not the analyzed program). Determining these positions is done by instrumenting the program analyzer and recording the relevant instructions affecting the branches the analyzer takes.

For example, for \figref{paexample} (a), we defined the analysis by the function $f_{overfit}$. For this function, only a subset of all AST nodes determine which of the three cases in the definition of $f_{overfit}$ will be used to compute the result of the analysis. Thus, we choose the modification position to be one of these AST nodes.

\subsection{Defining Relevant Program Modifications}\seclabel{modifications}
We now define two approaches for generating interesting program modifications that are potential counter-examples for the learned program analysis $\analysis$.

\subsubsection{Modification via Equivalence Modulo (EMA) Abstraction}\seclabel{ema}
The goal of EMA technique is to ensure that the candidate analysis $\analysis$ is robust to certain types of program transformations. To achieve this, we transform the statement at the selected program position in a semantically-preserving way, producing a~set of new programs. Moreover, while the transformation is semantic-preserving, it is also one that should not affect the result of the analysis $\analysis$.

More formally, an EMA transformation is a function $F_{ema} \colon \TL \times \mathbb{N} \rightarrow \powerset{(\TL)}$ which takes as input a program $p$ and a position in the program, and produces a set of programs that are a transformation of $p$ at position $n$. If the analysis $\analysis$ is correct, then these functions (transformations) have the following property:
\begin{equation}
\forall p' \in F_{ema}(p, n). \analysis(p) = \analysis(p')
\end{equation}

The intuition behind such transformations is to ensure stability by exploring \emph{local program modifications}. If the oracle detects the above property is violated, the current analysis $\analysis$ is incorrect and the counter-example program $p'$ is reported. Examples of applicable transformations are dead code insertion, variable names renaming or constant modification, although transformations to use can vary depending on the kind of analysis being learned.
For instance, inserting dead code that reuses existing program identifiers can affect flow-insensitive analysis, but should not affect a flow-sensitive analysis. The EMA property is similar to notion of algorithmic stability used in machine learning where the output of a~classifier should be stable under small perturbations of the input as well as the concept of equivalence modulo inputs used to validate compilers \cite{EmiTesting}.

\subsubsection{Modification via Global Jumps}\seclabel{gentest}
The previous modifications always generated semantic-preserving transformations. However, to ensure better generalization we are also interested in exploring changes to programs in $\programs_{\dataset}$ that may not be semantic preserving, defined via a function $F_{gj} \colon \TL \times \mathbb{N} \rightarrow \powerset{(\TL)}$. The goal is to discover a new program which exhibits behaviors not seen by any of the programs in $\programs_{\dataset}$ and is not considered by the currently learned analyzer~$\analysis$.

Overall, as shown in \secref{evaluation}, our approach for generating programs to test the analysis $\analysis$ via the functions $F_{gj}$ and $F_{ema}$ is an order of magnitude more efficient at finding counter-examples than naively modifying the programs in $\programs_{\dataset}$.

\section{Implementation and Evaluation}\seclabel{evaluation}
In this section we provide an implementation of our approach shown in~\figref{approach} as well as a detailed experimental evaluation instantiated to two challenging analysis problems for JavaScript: learning points-to analysis rules and learning allocation site rules. In our experiments, we show that:

\begin{itemize}
\item The approach can learn practical program analysis rules for tricky cases involving JavaScript's built-in objects. These rules can be incorporated into existing analyzers that currently handle such cases only partially.
\item The counter-example based learning is critical for ensuring that the learned analysis generalizes well and does not overfit to the training dataset.
\item Our oracle can effectively find counter-examples (orders of magnitude faster than random search).
\end{itemize}

These experiments were performed on a $28$ core machine with 2.60Ghz Intel(R) Xeon(R) CPU E5-2690 v4 CPU, running Ubuntu 16.04. In our implementation we parallelized both the learning and the search for the counter-examples. 

\paragraph{Training dataset}
We use the official ECMAScript (ECMA-262) conformance suite (\url{https://github.com/tc39/test262}) -- the largest and most comprehensive test suite available for JavaScript containing over 20 000 test cases.
As the suite also includes the latest version of the standard, all existing implementations typically support only a subset of the testcases. In particular, the \scode{NodeJS} interpreter v4.2.6 used in our evaluation can execute (\ie, does not throw a~syntax error) 15~675 tests which we use as the training dataset for learning.

\begin{table}[t]\centering
\caption{Program modifications used to instantiate the oracle (\secref{oracle}) that generates counter-examples for points-to analysis and allocation site analysis.}
{  \sffamily
\begin{tabular*}{0.9\textwidth}{@{}cc@{}}
\toprule

\multicolumn{2}{c}{~~\textbf{Program Modifications}}\\
\qquad \qquad \qquad \qquad ~~$F_{ema}$ \qquad \qquad \qquad \qquad & \qquad \qquad \qquad  \qquad $F_{gj}$ \qquad \qquad \qquad \qquad \\
\midrule
\noalign{\vskip 1mm}
~~Adding Dead Code & Adding Method Arguments\\
~~Renaming Variables & Adding Method Parameters\\
~~Renaming User Functions & Changing Constants \\
~~Side-Effect Free Expressions& \\
\bottomrule
\end{tabular*}
}

\tablabel{transformations}

\end{table} 

\paragraph{Program modifications}

We list the program modifications used to instantiate the oracle in \tabref{transformations}.
The semantic preserving program modifications that should not change the result of analyses considered in our work $F_{ema}$ are inserted dead code and renamed variables and user functions (together with the parameters) as well as generated expressions that are side-effect free (e.g, declaring new variables). Note that these mutations are very general and should apply to almost arbitrary property.
To explore new program behaviours by potentially changing program semantics we use program modifications $F_{gj}$ that change values of constants (strings and numbers), add methods arguments and add method parameters.

\subsection{Learning Points-to Analysis Rules for JavaScript}
We now evaluate the effectiveness of our approach for the task of learning a~points-to analysis for the JavaScript built-in APIs that affect the binding of \scode{this}. This is useful because existing analyzers currently either model this only partially \cite{GateKeeper,facebookflow} (\ie, cover only a subset of the behaviors of \scode{Function.prototype} APIs) or not at all \cite{Madsen:2013,Jang:2009}, resulting in potentially unsound results.

\begin{figure}[t]
\begin{tikzpicture}

\node at (0, 0) {
\footnotesize
\begin{minipage}{.45\textwidth}
\centering

\begin{lstlisting}[language=JavaScript, tabsize=2, escapechar=\&, basicstyle=\ttfamily\small]
global.length = 4;
var dat = [5, 3, 9, 1];
function isBig(value) {
  return value >= this.length;
}
\end{lstlisting}
\end{minipage}
};

\draw (2.3, -1.1) -- (2.3, 1.1);

\draw[thick] (-1.7, -0.8) -- (-0.8, -0.8);

\node at (6.8, 0) {
\footnotesize
\begin{minipage}{.66\textwidth}
\centering


\begin{lstlisting}[language=JavaScript, tabsize=2, escapechar=\&, basicstyle=\ttfamily\small]
// this points to global
dat.filter(isBig); // [5, 9]
// this points to boxed 42
dat.filter(isBig, 42); // []
// this points to dat object
dat.filter(isBig, dat); // [5, 9]
\end{lstlisting}
\end{minipage}
};

\end{tikzpicture}
\vspace{-2.9em}
\caption{JavaScript code snippet illustrating subset of different objects to which \scode{this} can point to depending on the context method \scode{isBig} is invoked in. 
}
\figlabel{this_binding}
\end{figure}
We illustrate some of the complexity for determining the objects to which \scode{this} points-to within the same method in \figref{this_binding}.
Here, \scode{this} points-to different objects depending on how the method is invoked and what values are passed in as arguments.
In addition to the values shown in the example, other values may be seen during runtime if other APIs are invoked, or the method \scode{isBig} is used as an object method or as a global method.

\begin{table}[t]\centering
\ra{0.98}
\caption{Dataset size, number of counter-examples found and the size of the learned points-to analysis for JavaScript APIs that affect the points-to set of \scode{this}.}
{  \sffamily
\begin{tabular*}{0.94\textwidth}{@{}lccccc@{}}
\toprule

~~Function Name&\quad Dataset Size \quad &\quad Counter-examples Found \quad & \quad Analysis Size$^*$ \quad \\
\midrule

~~\sffamily\scriptsize \textbf{Function.prototype}\\
~~~~\scode{call()} &  {\color{white}0}26 & 372 & 97 (18) \\
~~~~\scode{apply()} & {\color{white}00}6 & 182 & 54 (10) \\
~~\sffamily\scriptsize \textbf{Array.prototype}\\
~~~~\scode{map()} & 315 & {\color{white}0}64 & 36 (6)  \\
~~~~\scode{some()} & 229 & {\color{white}0}82 & 36 (6)\\
~~~~\scode{forEach()} & 604 & 177 & 35 (5)\\
~~~~\scode{every()} & 338 & {\color{white}0}31 & 36 (6)\\
~~~~\scode{filter()} & 408 & {\color{white}0}76 & 38 (6)\\
~~~~\scode{find()} & {\color{white}0}53 & {\color{white}0}73 & 36 (6)\\
~~~~\scode{findIndex()} & {\color{white}0}51 & {\color{white}0}96 & 28 (6) \\
~~\sffamily\scriptsize \textbf{Array}\\
~~~~\scode{from()} & {\color{white}0}32 & 160 & 57 (7) \\
~~\sffamily\scriptsize \textbf{JSON}\\
~~~~\scode{stringify()} & {\color{white}0}18 & {\color{white}0}55 & 9 (2)\\

\bottomrule

\multicolumn{4}{l}{\scriptsize $^*$ Number of instructions in \tlang{} (Number of \bcode{if} branches)}\\

\end{tabular*}
}
\tablabel{apis_eval}

\end{table}

\paragraph{Language \lang{}}
To learn points-to analysis, we use a domain-specific language \tlang{} with if statements (to synthesize branches for corner cases) and instructions to traverse the JavaScript AST in order to provide the specific analysis of each case. We provide a detailed list of the instructions with their semantics in \appref{DSLlanguage} and \appref{appendix_language}.

\paragraph{Learned analyzer}
A summary of our learned analyzer is shown in \tabref{apis_eval}. For each API we collected all its usages in the ECMA-262 conformance suite, ranging from only 6 to more than 600, and used them as initial training dataset for the learning. In all cases, a significant amount of counter-examples were needed to refine the analysis and prevent overfitting to the initial dataset.
On average, for each API, the learning finished in $14$ minutes, out of which $4$ minutes were used to synthesise the program analysis and $10$ minutes used in the search for counter-examples (cumulatively across all refinement iterations). The longest learning time was $57$ minutes for the \scode{Function.prototype.call} API for which we also learn the most complex analysis -- containing 97 instructions in \tlang{}.
We note that even though the APIs in \scode{Array.prototype} have very similar semantics, the learned programs vary slightly.
This is caused by the fact that different number and types of examples were available as the initial training dataset which means that also the oracle had to find different types of counter-examples.
We provide an example of the learned analysis in \appref{examples}. 

\subsection{Learning Allocation Site Analysis for JavaScript}
We also evaluate the effectiveness of our approach on a second analysis task -- learning allocation sites in JavaScript. This is an analysis that is used internally by many existing analyzers. The analysis computes which statements or expressions in a~given language result in an allocation of a~new heap object.

\begin{figure}[t]
\center
\begin{tikzpicture}

\node at (0, 0) {
\footnotesize
\begin{minipage}{.65\textwidth}
\centering

\begin{lstlisting}[language=JavaScript, tabsize=2, basicstyle=\ttfamily\small, keepspaces=true]
var obj = _{a: 7}_;
var arr = _[1, 2, 3, 4]_;
if (obj.a == _arr.slice(0,2)_) { ... }
var n = _new Number(7)_;
var obj2 = new Object(obj);
try {  ...  } catch (_err_) {  ...  }
\end{lstlisting}
\end{minipage}
};

\node[text width=6cm, align=center] at (6, 0) {\underline{Allocation Sites} \\(new object allocated)};


\end{tikzpicture}
\vspace{-2.5em}
\caption{Illustration of program locations (underlined) for which the allocation site analysis should report that a new object is allocated.}

\figlabel{alloc_example}
\end{figure}
We illustrate the expected output and some of the complexities of allocation site analysis on a example shown in \figref{alloc_example}.
In JavaScript, there are various ways how an object can be allocated including creating new object without calling a~constructor explicitly (for example by creating new array or object expression inline), creating new object by calling a~constructor explicitly using \scode{new}, creating a new object by calling a~method or new objects created by throwing an exception.
In addition, some of the cases might further depend on actual values passed as arguments. For example, calling a \scode{new Object(obj)} constructor with \scode{obj} as an argument does not create a new object but returns the \scode{obj} passed as argument instead.
The goal of the analysis is to determine all such program locations (as shown in \figref{alloc_example})  at which new object is allocated.

Consider the following simple, but unsound and imprecise allocation site analysis:
\[
\footnotesize
f_{alloc}(x) = \begin{cases}
true & \textbf{if~there is~~}\texttt{Argument:\scodevar{x}}\textbf{~or~}\texttt{NewExpression:\scodevar{x}} \\
false & \textbf{otherwise}
\end{cases}
\]
which states that a location $x$ is an allocation site if it is either an argument or a new expression. This analysis is imprecise because there are other ways to allocate an object (e.g., when creating arrays, strings, boxed values or by calling a function). It is also unsound, because the JavaScript compiler might not create a~new object even when \scode{NewExpression} is called (e.g., \scode{new Object(obj)} returns the same object as the given $obj$).

Instead of defining tricky corner cases by hand, we use our approach to learn this analyzer automatically from data. 
We instantiate the approach in a very similar way compared to learning points-to analysis by adjusting the language and how the labels in the training dataset are obtained (details provided in \appref{allocsite}).
For this task, we obtain $134~721$ input/output examples from the training data, which are further expanded with additional $905$ counter-examples found during 99 refinement iterations of the learning algorithm. For this (much higher than in the other analyzer) number of examples the synthesis time was $184$ minutes while the total time required to find counter-examples was $7$ hours.

The learned program is relatively complex and contains $135$ learned branches, including the tricky case where \scode{NewExpression} does not allocate a new object. 
Compared to the trivial, but wrong analysis $f_{alloc}$, the synthesized analysis marks over twice as many locations in the code as allocation sites ($\approx 21\text{K}$ vs $\approx 45\text{K}$). 

\subsection{Analysis Generalization}
We study how well the learned analyzer for points-to analysis works for unseen data. First, we manually inspected the learned analyzer at the first iteration of the $\textit{Synthesize}$ procedure (without any counter-examples generated). We did that to check if we overfit to the initial dataset and found that indeed, the initial analysis would \emph{not} generalize to some programs outside the provided dataset. This happened because the learned rules conditioned on unrelated regularities found in the data (such as variable names or fixed positions of certain function parameters). Our oracle, and the counter-example learning procedure, however, eliminate such kinds of non-semantic analyses by introducing additional function arguments and statements in the test cases.

Overfitting to the initial dataset was also caused by the large search space of possible programs in the DSL for the analysis. However, we decided not to restrict the language, because a more expressive language means more automation. Also, we did not need to provide upfront partial analysis in the form of a sketch~\cite{CombinatorialSketching}.

\paragraph{Oracle effectiveness for finding counter-examples}
We evaluate the effectiveness of our oracle to find counter-examples by comparing it to a random (``black box'') oracle that applies all possible modifications to a~randomly selected program from the training dataset.
For both oracles we measure the average number of programs explored before a counter-example is found and summarize the results in \tabref{oracle_eval}. In the table, we observe two cases: (i) early in the analysis loop when the analysis is imprecise and finding a counter-example is \emph{easy}, and (ii) later in the loop when \emph{hard} corner cases are not yet covered by the analysis. In both cases, our \emph{oracle guided by analysis} is orders of magnitude more efficient.

\begin{table}[t]\centering
\caption{The effect of using the learned analysis to guide the counter-example search.}
{  \sffamily
\begin{tabular*}{0.85\textwidth}{@{}lcc@{}}
\toprule

& \multicolumn{2}{c}{\textbf{Programs explored until first counter-example is found}}\\
~~Difficulty & \quad ``Black Box'' \quad & Guided by Analysis \\
\midrule

\noalign{\vskip 1mm}
~~Easy ($\approx 60\%$) & $146$ & $13$ \\
~~Hard ($\approx 40\%$) & $>3000$ & $130$\\
\bottomrule
\end{tabular*}
}

\tablabel{oracle_eval}
\end{table}

\paragraph{Is counter-example refinement loop needed?}
Finally, we compare the effect of learning with a refinement loop to learning with a~standard ``one-shot'' machine learning algorithm, but with more data provided up-front. For this experiment, we automatically generate a huge dataset $\dataset_{huge}$ by applying all possible program modifications (as defined by the oracle) on all programs in $\dataset$. For comparison, let the dataset obtained at the end of the counter-example based algorithm on $\dataset$ be $\dataset_{ce}$. The size of $\dataset_{ce}$ is two orders of magnitude smaller than $\dataset_{huge}$.

An analysis that generalizes well should be sound and precise on both datasets $\dataset_{ce}$ and $\dataset_{huge}$, but since we use one of the datasets for training, we use the other one to validate the resulting analyzer. For the analysis that is learned using counter-examples (from $\dataset_{ce}$), the precision is around $99.9\%$ with the remaining $0.01\%$ of results approximated to the top element in the lattice (that is, it does not produce a trivially correct, but useless result). However, evaluating the analysis learned from $\dataset_{huge}$ on $\dataset_{ce}$ has precision of only $70.1\%$ with the remaining $29.1\%$ of the cases being \emph{unsound}! This means that $\dataset_{ce}$ indeed contains interesting cases critical to analysis soundness and precision.

\paragraph{Summary}
Overall, our evaluation shows that the learning approach presented in our work can learn static analysis rules that handle various cases such as the ones that arise in JavaScript built-in APIs. The learned rules generalize to cases beyond the training data and can be inspected and integrated into existing static analyzers that miss some of these corner cases. We provide an example of both learned analyses in \appref{examples}. 

\section{Related Work}\seclabel{related}

\paragraph{Synthesis from examples}
Similar to our work, synthesis from examples typically starts with a domain-specific language (DSL) which captures a hypothesis space of possible programs together with a set of examples the program must satisfy and optionally an oracle to provide additional data points in the form of counter-examples using CEGIS-like techniques \cite{CombinatorialSketching}.
Examples of this direction include discovery of bit manipulation programs \cite{OracleSynthesis}, string processing in spreadsheets \cite{FlashFill}, functional programs \cite{FPSynthesis}, or data structure specifications~\cite{CommSynthesis}. A recent work has shown how to generalize the setting to large and noisy datasets \cite{NoiseLearning}.

Other recent works~\cite{Heule:2015,Jeon:2016} synthesize models for library code by collecting program traces which are then used as a specification. The key differences with our approach are that we (i) use large dataset covering hundreds of cases and (ii) we synthesize analysis that generalizes beyond the provided dataset.

\paragraph{Program analysis and machine learning}
Recently, several works tried to use machine learning in the domain of program analysis for task such as probabilistic type prediction \cite{BinaryTypes,JSNICE}, reducing the false positives of an analysis \cite{MayurAnalysis}, or as a~way to speed up the analysis \cite{BayesAnalysis,HeoOY16,ChaJO16} by learning various strategies used by the analysis.
A~key difference compared to our work is that we present a method to learn the static analysis rules which can then be applied in an iterative manner. This is a more complex task than \cite{BinaryTypes,JSNICE} which do not learn rules that can infer program specific properties and \cite{MayurAnalysis,BayesAnalysis,HeoOY16,ChaJO16} which assume the rules are already provided and typically learn a classifier on top of them.


\paragraph{Learning invariants}
In an orthogonal effort there has also been work on learning program invariants using dynamic executions. For recent representative examples of this direction, see \cite{InvLearningMadhu,TestsToProofs,SharmaInvariants}. The focus of all of these works is rather different: they work on a per-program basis, exercising the program, obtaining observations and finally attempting to learn the invariants.  Counter-example guided abstraction refinement (CEGAR) \cite{Clarke2000} is a classic approach for learning an abstraction (typically via refinement). Unlike our work, these approaches do not learn the actual program analysis and work on a per-program basis.

\paragraph{Scalable program analysis}
Another line of work considers scaling program analysis in hard to analyse domains such as JavaScript at the expense of analysis soundness \cite{Feldthaus:2013,Madsen:2013}.
These works are orthogonal to us and follow the traditional way of designing the static analysis components by hand, but in the future they can also benefit from automatically learned rules by techniques such as ours.




%

\ignore{
\paragraph{Program fuzzing}
Program fuzzing is a widely used ``black box'' technique for testing the robustness of wide range of programs including major browser implementations, compilers \cite{CSmith,EmiTesting,ManyCoreFuzzing}, SMT solvers \cite{Brummayer:2009:FDS} as well as to test static analyzers \cite{Cuoq:2012:TSA}.
The goal is to randomly generate new programs with the hope that they will lead to crashes or failing assertions.
The main challenge in these works is to design a program generator for a given domain such as C programs \cite{CSmith}, OpenCL programs \cite{ManyCoreFuzzing} or SMT formulas \cite{Brummayer:2009:FDS}.
In our work, instead of generating programs randomly we provide a technique for guiding the oracle by using the knowledge of the candidate program analysis under test.
}

\section{Conclusion and Future Work}
We presented a new approach for learning static analyzers from examples. Our approach takes as input a language for describing analysis rules, an abstraction function and an initial dataset of programs. Then, we introduce a counter-example guided search to iteratively add new programs that the learned analyzer should consider. These programs aim to capture corner cases of the programming language being analyzed. The counter-example search is made feasible thanks to an oracle able to quickly generate candidate example programs for the analyzer.

We implemented our approach and applied it to the setting of learning a~points-to and allocation site analysis for JavaScript. This is a very challenging problem for learning yet one that is of practical importance. We show that our learning approach was able to discover new analysis rules which cover corner cases missed by prior, manually crafted analyzers for JavaScript.

We believe this is an interesting research direction with several possible future work items including learning to model the interfaces of large libraries w.r.t to a given analysis, learning the rules for other analyzers (e.g., type analysis), or learning an analysis that is semantically similar to analysis written by hand.

\newpage
\bibliography{bib}

\begin{thebibliography}{10}

\bibitem{ChaJO16}
S.~Cha, S.~Jeong, and H.~Oh.
\newblock Learning a strategy for choosing widening thresholds from a large
  codebase.
\newblock In {\em Programming Languages and Systems - 14th Asian Symposium,
  {APLAS} 2016, Hanoi, Vietnam, November 21-23, 2016, Proceedings}, volume
  10017 of {\em Lecture Notes in Computer Science}, pages 25--41, 2016.

\bibitem{Clarke2000}
E.~Clarke, O.~Grumberg, S.~Jha, Y.~Lu, and H.~Veith.
\newblock {\em Counterexample-Guided Abstraction Refinement}, pages 154--169.
\newblock Springer Berlin Heidelberg, 2000.

\bibitem{Cousot:1977}
P.~Cousot and R.~Cousot.
\newblock Abstract interpretation: A unified lattice model for static analysis
  of programs by construction or approximation of fixpoints.
\newblock In {\em Proceedings of the 4th ACM SIGACT-SIGPLAN Symposium on
  Principles of Programming Languages}, pages 238--252, New York, NY, USA,
  1977. ACM.

\bibitem{facebookflow}
Facebook.
\newblock {Facebook} {Flow}: Static typechecker for javascript.
\newblock \url{https://github.com/facebook/flow}, 2016.

\bibitem{Feldthaus:2013}
A.~Feldthaus, M.~Sch\"{a}fer, M.~Sridharan, J.~Dolby, and F.~Tip.
\newblock Efficient construction of approximate call graphs for javascript ide
  services.
\newblock In {\em Proceedings of the 2013 International Conference on Software
  Engineering}, pages 752--761, 2013.

\bibitem{FPSynthesis}
J.~K. Feser, S.~Chaudhuri, and I.~Dillig.
\newblock Synthesizing data structure transformations from input-output
  examples.
\newblock In {\em Proceedings of the 36th {ACM} {SIGPLAN} Conference on
  Programming Language Design and Implementation, Portland, OR, USA, June
  15-17, 2015}, pages 229--239, 2015.

\bibitem{InvLearningMadhu}
P.~Garg, D.~Neider, P.~Madhusudan, and D.~Roth.
\newblock Learning invariants using decision trees and implication
  counterexamples.
\newblock In {\em Proceedings of the 43rd Annual {ACM} {SIGPLAN-SIGACT}
  Symposium on Principles of Programming Languages, {POPL} 2016}, pages
  499--512, 2016.

\bibitem{CommSynthesis}
T.~Gehr, D.~Dimitrov, and M.~T. Vechev.
\newblock Learning commutativity specifications.
\newblock In {\em Computer Aided Verification - 27th International Conference,
  {CAV} 2015, San Francisco, CA, USA, July 18-24, 2015, Proceedings, Part {I}},
  pages 307--323, 2015.

\bibitem{Giacobazzi:2015}
R.~Giacobazzi, F.~Logozzo, and F.~Ranzato.
\newblock Analyzing program analyses.
\newblock In {\em Proceedings of the 42Nd Annual ACM SIGPLAN-SIGACT Symposium
  on Principles of Programming Languages}, POPL '15, pages 261--273. ACM, 2015.

\bibitem{SAGE}
P.~Godefroid, M.~Y. Levin, and D.~Molnar.
\newblock Sage: Whitebox fuzzing for security testing.
\newblock {\em Queue}, 10(1):20:20--20:27, Jan. 2012.

\bibitem{GateKeeper}
S.~Guarnieri and B.~Livshits.
\newblock Gatekeeper: Mostly static enforcement of security and reliability
  policies for javascript code.
\newblock In {\em Proceedings of the 18th Conference on USENIX Security
  Symposium}, SSYM'09, pages 151--168, 2009.

\bibitem{FlashFill}
S.~Gulwani.
\newblock Automating string processing in spreadsheets using input-output
  examples.
\newblock In {\em Proceedings of the 38th {ACM} {SIGPLAN-SIGACT} Symposium on
  Principles of Programming Languages}, pages 317--330, 2011.

\bibitem{HeoOY16}
K.~Heo, H.~Oh, and H.~Yang.
\newblock Learning a variable-clustering strategy for octagon from labeled data
  generated by a static analysis.
\newblock In {\em Static Analysis - 23rd International Symposium, {SAS} 2016,
  Edinburgh, UK, September 8-10, 2016, Proceedings}, volume 9837 of {\em
  Lecture Notes in Computer Science}, pages 237--256. Springer, 2016.

\bibitem{Heule:2015}
S.~Heule, M.~Sridharan, and S.~Chandra.
\newblock Mimic: Computing models for opaque code.
\newblock In {\em Proceedings of the 2015 10th Joint Meeting on Foundations of
  Software Engineering}, ESEC/FSE 2015, pages 710--720, 2015.

\bibitem{Jang:2009}
D.~Jang and K.-M. Choe.
\newblock Points-to analysis for javascript.
\newblock In {\em Proceedings of the 2009 ACM Symposium on Applied Computing},
  SAC '09, pages 1930--1937, 2009.

\bibitem{TAJS}
S.~H. Jensen, A.~M{\o}ller, and P.~Thiemann.
\newblock Type analysis for javascript.
\newblock In {\em Proceedings of the 16th International Symposium on Static
  Analysis}, SAS '09, pages 238--255, Berlin, Heidelberg, 2009.
  Springer-Verlag.

\bibitem{Jeon:2016}
J.~Jeon, X.~Qiu, J.~Fetter-Degges, J.~S. Foster, and A.~Solar-Lezama.
\newblock Synthesizing framework models for symbolic execution.
\newblock In {\em Proceedings of the 38th International Conference on Software
  Engineering}, ICSE '16, pages 156--167, 2016.

\bibitem{OracleSynthesis}
S.~Jha, S.~Gulwani, S.~A. Seshia, and A.~Tiwari.
\newblock Oracle-guided component-based program synthesis.
\newblock In {\em Proceedings of the 32Nd ACM/IEEE International Conference on
  Software Engineering - Volume 1}, ICSE '10, pages 215--224, 2010.

\bibitem{BinaryTypes}
O.~Katz, R.~El{-}Yaniv, and E.~Yahav.
\newblock Estimating types in binaries using predictive modeling.
\newblock In {\em Proceedings of the 43rd Annual {ACM} {SIGPLAN-SIGACT}
  Symposium on Principles of Programming Languages, {POPL} 2016}, pages
  313--326, 2016.

\bibitem{TestsToProofs}
S.~Kowalewski and A.~Philippou, editors.
\newblock {\em Tools and Algorithms for the Construction and Analysis of
  Systems, {TACAS} 2009}, volume 5505 of {\em Lecture Notes in Computer
  Science}. Springer, 2009.

\bibitem{EmiTesting}
V.~Le, M.~Afshari, and Z.~Su.
\newblock Compiler validation via equivalence modulo inputs.
\newblock In {\em Proceedings of the 35th ACM SIGPLAN Conference on Programming
  Language Design and Implementation}, PLDI '14, pages 216--226, 2014.

\bibitem{Livshits:2015}
B.~Livshits, M.~Sridharan, Y.~Smaragdakis, O.~Lhot\'{a}k, J.~N. Amaral,
  B.-Y.~E. Chang, S.~Z. Guyer, U.~P. Khedker, A.~M{\o}ller, and D.~Vardoulakis.
\newblock In defense of soundiness: A manifesto.
\newblock {\em Commun. ACM}, 58(2):44--46, Jan. 2015.

\bibitem{StatisticalLearningTheory}
U.~v. Luxburg and B.~Schoelkopf.
\newblock Statistical learning theory: Models, concepts, and results.
\newblock In {\em Inductive Logic}, pages 651--706. 2011.

\bibitem{Madsen:2013}
M.~Madsen, B.~Livshits, and M.~Fanning.
\newblock Practical static analysis of javascript applications in the presence
  of frameworks and libraries.
\newblock In {\em Proceedings of the 2013 9th Joint Meeting on Foundations of
  Software Engineering}, ESEC/FSE 2013, pages 499--509, New York, NY, USA,
  2013. ACM.

\bibitem{MayurAnalysis}
R.~Mangal, X.~Zhang, A.~V. Nori, and M.~Naik.
\newblock A user-guided approach to program analysis.
\newblock In {\em Proceedings of the 2015 10th Joint Meeting on Foundations of
  Software Engineering, {ESEC/FSE} 2015}, pages 462--473, 2015.

\bibitem{BayesAnalysis}
H.~Oh, H.~Yang, and K.~Yi.
\newblock Learning a strategy for adapting a program analysis via bayesian
  optimisation.
\newblock In {\em Proceedings of the 2015 {ACM} {SIGPLAN} International
  Conference on Object-Oriented Programming, Systems, Languages, and
  Applications, {OOPSLA} 2015}, pages 572--588, 2015.

\bibitem{ID3}
J.~R. Quinlan.
\newblock Induction of decision trees.
\newblock {\em Mach. Learn.}, 1(1):81--106, Mar. 1986.

\bibitem{NoiseLearning}
V.~Raychev, P.~Bielik, M.~Vechev, and A.~Krause.
\newblock Learning programs from noisy data.
\newblock In {\em Proceedings of the 43rd Annual ACM SIGPLAN-SIGACT Symposium
  on Principles of Programming Languages}, POPL '16, pages 761--774, 2016.

\bibitem{JSNICE}
V.~Raychev, M.~Vechev, and A.~Krause.
\newblock Predicting program properties from "big code".
\newblock In {\em Proceedings of the 42Nd Annual ACM SIGPLAN-SIGACT Symposium
  on Principles of Programming Languages}, POPL '15, pages 111--124, 2015.

\bibitem{SharmaInvariants}
R.~Sharma, S.~Gupta, B.~Hariharan, A.~Aiken, and A.~V. Nori.
\newblock Verification as learning geometric concepts.
\newblock In {\em Static Analysis - 20th International Symposium, {SAS} 2013,
  Seattle, WA, USA, June 20-22, 2013. Proceedings}, pages 388--411, 2013.

\bibitem{PGL-014}
Y.~Smaragdakis and G.~Balatsouras.
\newblock Pointer analysis.
\newblock {\em Foundations and Trends® in Programming Languages}, 2(1):1--69,
  2015.

\bibitem{CombinatorialSketching}
A.~Solar{-}Lezama, L.~Tancau, R.~Bod{\'{\i}}k, S.~A. Seshia, and V.~A.
  Saraswat.
\newblock Combinatorial sketching for finite programs.
\newblock In {\em Proceedings of the 12th International Conference on
  Architectural Support for Programming Languages and Operating Systems,
  {ASPLOS} 2006}, pages 404--415, 2006.

\end{thebibliography}
\bibliographystyle{abbrv}

\newpage
\appendix
\section*{Appendix}
Here we provide a detailed description of how we instantiated the learning approach presented in our work to the tasks of learning points-to and allocation site analysis for JavaScript. In particular, the appendix contains following sections:

\begin{enumerate}[label=\scode{\Alph*:}, itemindent=3em,leftmargin=2em]
\item Instantiation of the learning points-to analysis
\item Description of \tlang{} language for points-to analysis
\item Formal semantics of \tlang{} language
\item Instantiation of the learning for allocation site analysis
\item Examples of learned program analyses
\item Implementation details of our approach
\end{enumerate}

\section{Points-to Analysis}\seclabel{pointsto}
In this section we present an instantiation of our approach to the task of learning transformers/rules for points-to analysis.
%
The goal of points-to analysis is to answer queries of the type $q \colon V \rightarrow \powerset{(H)}$, where $V$ is a set of program variables and $H$ is a heap abstraction (\eg, allocation sites).
That is, the goal is to compute the set of (abstract) objects to which a variable may point-to at runtime.
Similar to the example illustrated in \secref{overview}, to answer such queries a common line of work \cite{PGL-014,GateKeeper,Madsen:2013} uses a declarative approach where the program is abstracted as a set of facts and the analysis is defined declaratively (\eg, as a set of Datalog rules) using inference rules that are applied until a fixed point is reached.


%

\paragraph{Our goal}
Our goal is to learn the inference rules that define the analysis, from data, as described in our approach so far. In particular, we would like to infer rules of the following general shape:
\[
\footnotesize
\inferrule*[Right=\footnotesize \textsc{$~$[General]}]
{\scode{VarPointsTo(}v_2, h\scode{)} \+ v_2 = f(v_1)}
{\scode{VarPointsTo(}v_1, h\scode{)}}
\]
where the goal of learning is to find a set of functions $f$ that, when used in the points-to analysis, produce precise results (as defined earlier). However, we focus our attention not on learning the standard and easy to define rules, as the one for assignment, but on rules that are hard and tricky to model by hand and are missed by existing analyzers. In particular, consider the following subset of inference rules that capture the points-to sets for the \scode{this} variable in JavaScript. This rule has the following shape:
\[
\footnotesize
\inferrule*[Right=\footnotesize \textsc{$~$[This]}]
{\scode{VarPointsTo(}v_2, h\scode{)} \+ v_2 = f(\scode{this})}
{\scode{VarPointsTo(this}, h\scode{)}}
\]
which is an instantiation of the general rule for the \scode{this} variable by setting $v_1 = \scode{this}$.
In JavaScript, designing such rules is a challenging task as there are many corner cases and describing those precisely requires more inference rules than the rest of the (standard) analysis rules.
Further, because assigning a value to the \scode{this} object is not allowed (\ie, using \scode{this} as a left-hand side of an assignment expression), the value of \scode{this} at runtime is not observed at the program level, yet assignments do occur internally in the interpreter and the runtime.
Complicating matters, the actual values of the \scode{this} reference can depend on the particular version of the interpreter.




\subsection{Instantiating our Learning Approach}
We now define the necessary components required to instantiate the learning approach described so far. Most of the instantiations are fairly direct except for the language $\lang$, described separately in \appref{DSLlanguage}.

\tikzset{
    >=stealth',
    pil/.style={
           thin,
           shorten <=2pt,
           shorten >=2pt,}
}

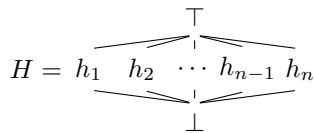
\begin{wrapfigure}[10]{r}{0.35\textwidth}
\centering
\vspace{-1.75em}
\begin{tikzpicture}
\node[] (bottom) {$\bot$};
\node[] (top) at ([yshift=40pt]bottom) {$\top$};

\node[] (h1) at ([xshift=-40pt, yshift=20pt]bottom) {$h_1$};
\node[] (h2) at ([xshift=-20pt, yshift=20pt]bottom) {$h_2$};
\node[] (h3) at ([yshift=20pt]bottom) {$\cdots$};
\node[] (h4) at ([xshift=20pt, yshift=20pt]bottom) {$h_{n-1}$};
\node[] (h5) at ([xshift=40pt, yshift=20pt]bottom) {$h_{n}$};

\node[] at ([xshift=-20]h1) {$H =$};

\draw[pil] (bottom.north) -- (h1.south);
\draw[pil] (bottom.north) -- (h2.south);
\draw[pil] (bottom.north) -- (h3.south);
\draw[pil] (bottom.north) -- (h4.south);
\draw[pil] (bottom.north) -- (h5.south);

\draw[pil] (h1.north) -- (top.south);
\draw[pil] (h2.north) -- (top.south);
\draw[pil] (h3.north) -- (top.south);
\draw[pil] (h4.north) -- (top.south);
\draw[pil] (h5.north) -- (top.south);
 
\end{tikzpicture}
\caption{Lattice of context-insensitive abstract heap locations $H$ for points-to analysis.}
\figlabel{lattice}
\end{wrapfigure}
\paragraph{Lattice of abstract heap locations}
\figref{lattice} shows the lattice $(\mathcal{H}, \sqsubseteq)$ used to represent the abstract domain of heap locations $H$.
The abstraction function $\alpha \colon O \rightarrow H$ maps the concrete objects seen at runtime to abstract heap locations represented using a context-insensitive allocation site abstraction~$H$.
The lattice is quite simple and consists of the standard elements $\top$, $\bot$ and elements corresponding to individual heap locations $h_1 \cdots h_n$ that are not comparable.

\paragraph{Concrete and abstract program semantics}

The concrete properties we are tracking and their abstract counterpart as described in \secref{analysisCorrectness} are instantiated by setting $\mathcal{C} := O$, $\mathcal{A} := H$ and $\mathbb{N} := \langle V, \mathbb{I}^{*} \rangle$.
That is, all concrete program behaviors are captured by a function $\ConcSem{p} \colon \langle V, \mathbb{I}^{*} \rangle \rightarrow \powerset{(O)}$ that for each program variable $V$ sensitive to the \textit{k}-most recent call sites $I$ computes a set of possible concrete objects seen at runtime $O$.
The abstract semantics are similar except that we instantiate the abstract domain to be the lattice describing heap-allocated objects $H$.
We discuss how we obtain the concrete behaviors $\ConcSem{p}_{ti}$ after running the program on a set of test inputs $ti$ in \appref{obtainingTrace}.






\begin{table}[t]\centering
\caption{Program modifications used to instantiate the oracle (\secref{oracle}) that generates counter-examples for points-to analysis and allocation site analysis.}
{  \sffamily
\begin{tabular*}{0.9\textwidth}{@{}cc@{}}
\toprule

\multicolumn{2}{c}{~~\textbf{Program Modifications}}\\
\qquad \qquad \qquad \qquad ~~$F_{ema}$ \qquad \qquad \qquad \qquad & \qquad \qquad \qquad  \qquad $F_{gj}$ \qquad \qquad \qquad \qquad \\
\midrule
\noalign{\vskip 1mm}
~~Adding Dead Code & Adding Method Arguments\\
~~Renaming Variables & Adding Method Parameters\\
~~Renaming User Functions & Changing Constants \\
~~Side-Effect Free Expressions& \\
\bottomrule
\end{tabular*}
}

\tablabel{transformations}

\end{table} 

\paragraph{Program modifications}

We list the program modifications used to instantiate the oracle in \tabref{transformations}.
The semantic preserving program modifications that should not change the result of points-to analysis $F_{ema}$ are inserted dead code and renamed variables and user functions (together with the parameters) as well as generated expressions that are side-effect free (e.g, declaring new variables).
To explore new program behaviours by potentially changing program semantics we use program modifications $F_{gj}$ that change values of constants (strings and numbers), add methods arguments and add method parameters.



\ignore {

\input{figures/points_to_domains}
\subsection{Points-to Analysis Formalization}

We now provide a formal description of the points-to analysis of \scode{this} keyword considered in our work.
In general our work follows the idea of running a static analysis of an input program that maintains the state of the environment, heap, and various other instrumentation \cite{Liang:2010} \TODO{add more citations}.
Such analyses are typically defined over an abstract program semantics that apply abstraction to the concrete program semantics at various places during the program execution, such as the classical object allocation site abstraction.
However, as discussed in \secref{point_to_overview}, since the standard language statements are not relevant for the analysis (the assignment to \scode{this} is not directly observable in the program) we do not follow this approach and do not provide abstract syntax and abstract semantics of the standard language statements.
Instead, we define the program analysis using concrete program semantics that precisely capture the real program executions (as we will see in \secref{?}).
This allows us to learn programs analyzers without having to define abstract program semantics as we can obtain concrete execution traces by simply running the programs.
We note that the same approach is still applicable if abstract program semantics are defined.


\paragraph{Syntactic and Semantic Domains}

The relevant set of syntactic and semantic domains for the analysis are shown in \figref{this_domains}.
The program is represented as a set of nodes $\mathbb{N}$ that constitute the abstract syntax tree (AST) corresponding to the program.
This is an alternative representation of programs compared to the traditional approach of representing programs as a set of programs points that correspond to the program statements.
Note however that these two representations are interchangeable as we can can describe any statement using its AST representation and vice versa.
Here we omit description of other syntactic domains such as fields, local variables or array indexes as these are not relevant for determining the location to which \scode{this} variable points-to.

Given syntactic domain that represents program as an AST we now describe program execution by defining relevant semantic domains.
The set $\mathbb{O}$ denotes set of all concrete objects in the program, the set $\mathbb{T}$ denotes set of all references to \scode{this} variable in the program and a set of call sites $\mathbb{I}$ in the program.
Note that in practice both sets of references to \scode{this} variable and call sites can be represented as a mapping to the corresponding AST node where \scode{this} variable is used and call site is located respectively.

\paragraph{Points-to Queries}
Finally, the goal of the points-to analysis is to answer the query $q$ that given a \scode{this} variable and a current execution call stack returns the concrete object to which \scode{this} variable points-to, $null$ or $undefined$.
In general the points-to analysis might not be able to always precisely identify a single object to which a reference points to (\eg, because of aliasing or the fact that its flow or path insensitive) and therefore the query usually returns a set of objects.
However, as we will see in \secref{?}, for the points-to analysis of \scode{this} variable in JavaScript we can achieve precise results and therefore the analysis we are interested in synthesising will return only a single object.

\subsection{Designing Points-to Analysis}

Having formally defined what is the expected output of points-to analysis, we now turn our attention to designing a program analysis that can be used to answer such queries.
In particular, we first describe how can we represent the semantic domains from \figref{this_domains} and then define a language called \TODO{??} that is used to describe a space of possible points-to analyses. The goal of learning described in \secref{?} will then be to find a analysis from this hypothesis space that is precise at answering points-to queries.

\paragraph{Representing Semantic Domains}
As a first step we define how to represent semantic domains for concrete objects $\mathbb{O}$, this variable $\mathbb{T}$ and call sites $\mathbb{I}$ as the analysis will need to refer to the element of these sets.
For this purpose we execute the program using concrete or abstract semantics and relate observed values of each element from $\mathbb{O}, \mathbb{T}$ and $\mathbb{I}$ to an element from syntactic domain $\mathbb{N}$.
That is, whenever the execution observes a concrete value $o$ or \scode{this} reference $t$ it is related to its corresponding AST node $n$. Similarly the call sites $i$ simply point to the AST node containing the statement with the call.
Using this approach we can define program execution trace:

\begin{definition}{\textit{Program Trace}}\deflabel{program_trace}
A program execution trace is a sequence of tuples in the form $\langle (\mathbb{O} \cup \mathbb{T} \cup \mathbb{I}) \times{} \mathbb{N} \rangle$, where each element corresponds to either reading a value of concrete object, a value of \scode{this} reference or a call site together with corresponding location in the AST.
\end{definition}

We note that this approach is suitable for executing programs with concrete as well as abstract semantics as they both simply relate the observed values to the nodes in the AST.
The only difference is that the observed values are different with concrete semantics (\eg, the concrete objects $o \in \mathbb{O}$) compared to those obtained using abstract semantics (\eg, allocation sites where the abstraction function $\alpha$ maps each concrete object $o$ to an allocation site denoting position in the program where $o$ was allocated).


\textit{Example}
\TODO{Snippet of code, its corresponding AST and the execution trace illustrating the mapping between the two.}

%



\paragraph{Designing Hypothesis Space of Points-to Analyses}
Our goal now is to design a program analysis $A_{pt}$ that can precisely answer the points-to queries of the form: 

\[
A_{pt}: \mathbb{T} \times{} \mathbb{I}^* \rightarrow (\mathbb{O} \cup \{null\} \cup \{undefined\})
\]

that is, given a \scode{this} reference and a call stack the analysis should return the object to which \scode{this} points-to or one of special $null$ or $undefined$ values.
In particular, what we want to define is not the analysis itself but instead a hypothesis space that contains many analysis capable of answering queries of this form.
The individual program analyses in this hypothesis space will naturally vary in the precision they achieve in practice and the goal of learning is then to find an analysis that is best at explaining the data.
As we will see later in \secref{?} and \secref{?} the hypothesis space might contain several analysis that all classify the training programs perfectly (in which case the learning pick the analysis it believes generalizes best) as well as none (in which case the goal of learning is to find the analysis with the least amount of errors).

In order to represent the hypothesis space of points-to analyses we use the following key idea -- we can uniquely determine the points-to result by selecting one or more positions in the program execution trace. The object values seen at these positions then form the points-to set for that given query (as illustrated by the arrows in \figref{this_binding}). Furthermore, since the program execution trace contains a mapping to the AST (see \defref{program_trace}) we can alternatively answer such queries by selecting one or more positions in the AST.

Using this idea we can now concisely represent the hypothesis space of the analysis $A_{pt}$ by defining a following domain specific language called \tlang{}.
As we include information about the call trace such analysis is context-sensitive while at the same time path and flow-insensitive which is sufficient abstraction that allows us to answer majority of the queries precisely.


%

%
%
%

%
%
%
%
%

%
%

%

%
}

\ignore{
\subsection{Challenges}\seclabel{challenges}

Answering points-to queries for \scode{this} keyword is especially challenging for a language like JavaScript because:
i) the traditional points-to algorithms do not apply as assigning a value to the \scode{this} object (\ie, as a left-hand side of an assignment expression) is an invalid operation, and ii) the operations that determine the value of \scode{this} observed at runtime are usually implemented within the JavaScript interpreter and are not directly observable to the analysis (\ie, an assignment to \scode{this} is never observed in the program).

To illustrate some of the complexity of determining the objects to which \scode{this} reference points to we discuss the most commonly used contexts where \scode{this} is used.
Various contexts of using \scode{this} keyword in JavaScript language are shown in \figref{this_binding} and include:

\begin{itemize}
\item \textbf{\figref{this_binding} a: Global Context}. When used in global context (\ie, outside of any function), then the object to which \scode{this} points-to depends on the JavaScript environment used. For example, within a web browser \scode{this} points to the \scode{window} object whereas in \scode{Node.js} it points-to the \scode{global} object.

\item \textbf{\figref{this_binding} b: Constructor Context}. When a function is used as a constructor (with the \scode{new} keyword), then \scode{this} points-to the new object that is being constructed.

\item \textbf{\figref{this_binding} c: Object Context}. When a function is not used as a constructor but called as an object method, then \scode{this} points-to the object the method is called on.

\item \textbf{\figref{this_binding} d: DOM Handlers}. When a function is used as a callback handler on a DOM object, then when invoked, \scode{this} points-to the object on which the callback handler was registered on. Note however that this behaviour might not be consistent among all the browsers as for example Internet Explorer 8.0 sets \scode{this} such that it points-to the global \scode{window} object.

\item \textbf{\figref{this_binding} e: Explicit Binding}.
A common pitfall in JavaScript is that \scode{this} keyword does not follow the rules of lexical scope (in contrast to variables).
As a result, when defining an inner function one needs to be careful when using \scode{this} as it might point-to a different object than expected.
For example, if anonymous function used in \scode{setTimeout} (in \figref{this_binding} e) would not use \scode{bind} statement, then \scode{this} would point-to the global \scode{window} object instead of the \scode{btn}.
To address this issue one can explicitly bind value to which \scode{this} object points-to within a function using one of the methods provided by the framework such as \scode{bind}, \scode{call} or \scode{apply}.

\item \textbf{\figref{this_binding} f: Library Binding}.
Finally, alike to the binding value if \scode{this} keyword within a function defined by the framework, some libraries provide similar functionality by supporting binding of the callbacks to a specified value (passed as an optional argument).
For example, second argument of \scode{Array.prototype.forEach} determines the value \scode{this} is bound to when executing a callback for each element of the input array.
Similarly, when using \scode{each} method defined by \scode{Underscore.js} library \footnote{http://underscorejs.org/} the value of \scode{this} is bound to the third argument.

\end{itemize}

As can be seen the values bound to the \scode{this} keyword are often determined dynamically based on how given function is invoked (e.g., object method, global method, bound method) which makes the actual value difficult to predict even by an experienced developers which can lead to subtle mistakes such as the one shown in \figref{this_binding} e.
To make things even worse, the actual values of \scode{this} reference can depend on the particular interpreter and can be affected by hard to analyze libraries (e.g., \scode{jQuery} or \scode{Underscore.js}) as well as future changes to the JavaScript language (e.g., Arrow functions introduced in 6th Edition of the ECMAScript standard \footnote{http://www.ecma-international.org/ecma-262/6.0/}).
Therefore, in our work we explore a new and compelling direction that given a particular interpreter, environment and a set to programs learns such analysis directly from the data.
}

\section{Language for Points-To Inference Rules}\seclabel{DSLlanguage}

We now provide a definition of our domain specific language \tlang{}, an instantiation of the template language $\mathcal{L}$ shown in \figref{template}. Our main goal was to design a~language \tlang{} that is fairly generic: (i) it does not require the designer to provide specific knowledge about the analysis rules, and (ii) the language can be used to describe rules beyond those of points-to analysis. Point (i) is especially important as specifying tricky parts of the analysis rules by hand requires substantial effort, which is exactly the process we would like to automate. Indeed, we aim at a language that is expressive enough to capture complex rules which use information from method arguments, fields, assignments, etc., yet can be automatically discovered during the learning.

To achieve this, the main idea is to define \tlang{} to work over Abstract Syntax Tree (AST) by providing means of navigating and conditioning on different parts of the tree. Further, we do not require the analysis to compute the results directly (e.g., a concrete points-to set for a given location). Instead, we allow the results to be specified indirectly by means of navigating to an AST position that determines the result. For example, such locations in the AST correspond to program positions with the same points-to set for points-to analysis, or to declaration sites for scope analysis or to program positions with the same type for type analysis. Next, we discuss the syntax and semantics of \tlang{}.





\paragraph{Syntax}

The syntax of \tlang{} is summarized in \figref{language} and consists of two kinds of basic instructions -- \scode{Move} instructions that navigate over the tree and \scode{Write} instructions that accumulate facts about the visited nodes.
We split the \scode{Move} instructions into three groups where \scode{Move$_{\scode{core}}$} include language and analysis independent instructions that navigate over trees, \scode{Move$_{\scode{js}}$} include instructions that navigate to a set of interesting program locations that are specific to the JavaScript language. Finally we include \scode{Move$_{\scode{call}}$} which allows learning of a call-site sensitive analysis. Using the \scode{Move} and \scode{Write} instructions we then define an action to be a sequence of \scode{Move} instructions and a guard to be a sequence of both \scode{Move} and \scode{Write} instructions.


\begin{figure}[t]
{

\begin{align*}
m \in \scode{Move}_{\scode{core}} ::=&~ \scode{{\color{red!50!black}Up} | {\color{red!50!black}Left} | {\color{red!50!black}Right} | {\color{red!50!black}DownFirst} | {\color{red!50!black}DownLast} | {\color{red!50!black}Top}}\\[1pt]
m \in \scode{Move}_{\scode{js~~}} ::=&~ \scode{{\color{red!50!black}GoToGlobal} | {\color{red!50!black}GoToUndef} | {\color{red!50!black}GoToNull} | {\color{red!50!black}GoToThis} | {\color{red!50!black}UpUntilFunc} }\\[1pt]
m \in \scode{Move} ::=&~ \scode{Move}_{\scode{core}}~\cup~\scode{Move}_{\scode{call}}~\cup~\scode{Move}_{\scode{js} } \quad \quad \quad ~ m \in \scode{Move}_{\scode{call}} ::=~\scode{{\color{red!50!black}GoToCaller} }\\[1pt]
w \in \scode{Write} ::=&~ \scode{{\color{blue!50!black}WriteValue}\,|\,{\color{blue!50!black}WritePos}\,|\,{\color{blue!50!black}WriteType}\,|\,{\color{blue!50!black}HasLeft}\,|\,{\color{blue!50!black}HasRight}\,|\,{\color{blue!50!black}HasChild} } 
\end{align*}
\vspace{-2.2em}
\begin{align*}
a \in Actions_{pt} ::=&~ \epsilon \mid \scode{Move}~~\texttt{;}~~a\\
g \in Guards_{pt} ::=&~ \epsilon \mid \scode{Move}~~\texttt{;}~~g \mid \scode{Write}~~\texttt{;}~~g\\
ctx \in Context  ::=&~ (N \cup \Sigma \cup \mathbb{N})^* \\[2pt]
l \in \lang_{pt} ::=&~ \epsilon \mid a \mid \bcode{if}~~g = ctx~~\bcode{then}~~l ~~\bcode{else}~~l
\end{align*}
%
}
\vspace{-1.6em}
\caption{Language \tlang{} for expressing the result of points-to query by means of navigating over an abstract syntax tree.}
\figlabel{language}
\end{figure}

\paragraph{Semantics}

Programs from \tlang{} operate on a state $\sigma$ defined as follows: $\sigma = \langle \tree, \node, ctx, i\rangle \in States$ where the domain $States = AST \times X \times Context \times \mathbb{I}^*$.
In a state $\sigma = \langle \tree, \node, ctx, i\rangle$, $\tree$ is an abstract syntax tree, $\node$ is the current position in the tree, $ctx$ is the currently accumulated context and $i$ is the current call trace.
The accumulated context $ctx \in Context = (N \cup \Sigma \cup \nats)^{*}$ by a \tlang{} program is a sequence of observations on the tree where each observation can be a non-terminal symbol $N$ from the tree, a terminal symbol $\Sigma$ from the tree or a natural number in $\nats$.
Initially, execution starts with the empty context $[] \in Context$ and the AST $t$, initial node $n$ and current call trace $i$ supplied as arguments.


For a program $p \in \tlang{}$, a tree $\tree \in AST$, and a position $\node \in X$ in the tree, we say that program $p$ computes a position $\node' \in X$, denoted as $p(\tree, \node, i) = \node'$, iff there exists a~sequence of transitions from $\langle p, \tree, \node, [], i \rangle$ to $\langle \epsilon, \tree, \node', [], i \rangle$. That is, $\node'$ is the last visited position by executing the program $p$ on a tree $t$ starting at position $\node$.
The context is empty both at the beginning and at the end of execution as it is used only to evaluate the \bcode{if} condition when deciding which branch to take.
%
We provide the small-step semantics of \scode{Move} and \scode{Write} instructions as well as the \bcode{if-then-else} statement, in the \appref{appendix_language}.

%
%

\paragraph{Example}
Consider the following program in \tlang{} that encodes the \textsc{Assign} rule illustrated in \secref{overview}:

{
\[
\small
f(t, \node, i) = \begin{cases}
\scode{\red{Right}} & \textbf{if~}\scode{\blue{WritePos} \red{Up} \blue{WriteType} = 1 Assignment} \\
\scode{\red{Top}} & \textbf{else}
\end{cases}
\]
}
\normalsize

When executed on a tree $t$ in \figref{paexample} (b) starting at position $n$ = \scode{Identifier:}\scodevar{a}, the program $f$ first executes the guard \scode{\blue{WritePos} \red{Up} \blue{WriteType}} which starts by writing value \scode{1} as the node at current position is the first child, then navigates to the position of parent node and writes its type \scode{Assignment}. This collected context \scode{1 Assignment} is then compared to the one specified in the condition. The equality is satisfied and the program takes the \bcode{if} branch, resets the current position back to the position $\node$ before executing the context inside the \bcode{if} branch, and then continues executing the code inside the \bcode{if}, \scode{\red{Right}}, which navigates to its right sibling. This sibling is also the output of executing program $f(t, \node, i)$.


\section{Formal Semantics of $\tlang{}$ Language}\seclabel{appendix_language}
Here, we provide the semantics of all \scode{Move} and \scode{Write} instructions as presented in \appref{DSLlanguage}. Further, we provide small-step semantics of \tlang{} language.

\subsection{\tlang{}: Semantics of Instructions}\seclabel{flatprograms}

The semantics of the write instructions are described by the \textsc{[Write]} rule in \figref{tlangsemantics}. Each write accumulates a value $c$ to the context $ctx$ according to the function $wr$: 
\[
wr \colon \scode{Write} \times AST \times X \times \mathbb{I}^* \rightarrow N \cup \Sigma \cup \nats
\]
defined as follows:

\begin{itemize}
\item $wr(\scode{\color{blue!50!black}WriteType}, \tree, \node, i)$ returns $x$ where $x \in N$ is the non-terminal symbol at node $\node$.
\item $wr(\scode{\color{blue!50!black}WriteValue}, \tree, \node, i)$ returns the terminal symbol at node $\node$ if one is available or a special value $0$ otherwise, and
\item $wr(\scode{\color{blue!50!black}WritePos}, \tree, \node, i)$ returns a number $x \in \nats$ that is the index of $\node$ in the list of children kept by the parent of $\node$.
\item $wr(\scode{\color{blue!50!black}HasLeft}, \tree, \node, i)$ and $wr(\scode{\color{blue!50!black}HasRight}, \tree, \node, i)$ return $1$ if the $\node$ has a left (right) sibling and~$0$ otherwise. 
\item $wr(\scode{\color{blue!50!black}HasChild}, \tree, \node, i)$ returns $1$ if the $\node$ has atleast one children and $0$ otherwise.
\item $wr(\scode{\color{blue!50!black}HasCaller}, \tree, \node, i)$ returns $1$ if the call trace is non-empty (\ie, $\abs{i} > 0$) and $0$ otherwise.

\end{itemize}

\begin{figure}[t]
\centering
\begin{tabular}{c}
\footnotesize
\begin{minipage}{3.8in}
\[
\begin{array}{c}
\inferrule*[Right=]
{\quad \+ \quad \+ \tree \in AST \+ \node \in X \+ ctx \in Context \+ i \in \mathbb{I}^* \+ \statements \in \scode{\tlang{}}}
{}
\\[5pt]
\inferrule*[Right=\footnotesize \textsc{$~$[Move]}]
{op \in \scode{Move} \+ \node' \times i' = mv(op, \tree, \node, i) \+ \node' \notin \{\bottom, \top\}}
{\langle op::\statements, \tree, \node, ctx, i \rangle \xrightarrow{} \langle \statements, \tree, \node', ctx, i' \rangle}
\\ \\
\inferrule*[Right=\footnotesize \textsc{$~$[Move-Fail]}]
{op \in \scode{Move} \+ \node' \times i' = mv(op, \tree, \node, i) \+ \node' \in \{\bottom, \top\}}
{\langle op::\statements, \tree, \node, ctx, i \rangle \xrightarrow{} \langle \epsilon, \tree, \node', ctx, i \rangle}
\\ \\
\inferrule*[Right=\footnotesize \textsc{$~$[Write]}]
{op \in \scode{Write} \+ c = wr(op, \tree, \node, i)}
{\langle op::\statements, \tree, \node, ctx, i \rangle \xrightarrow{} \langle \statements, \tree, \node, ctx \cdot c, i \rangle}
\\ \\
\inferrule*[Right=\footnotesize \textsc{$~$[If-True]}]
{op \in \bcode{if}~~g = ctx~~\bcode{then}~~l_{true} ~~\bcode{else}~~l_{false} \+
\langle g, \tree, \node, [], i \rangle \rightarrow \langle \epsilon, \tree', \node', ctx', i' \rangle \+
ctx = ctx'
}
{\langle op, \tree, \node, ctx, i \rangle \xrightarrow{} \langle l_{true}, \tree, \node, ctx, i \rangle}
\\ \\
\inferrule*[Right=\footnotesize \textsc{$~$[If-False]}]
{op \in \bcode{if}~~g = ctx~~\bcode{then}~~l_{true} ~~\bcode{else}~~l_{false} \+
\langle g, \tree, \node, [], i \rangle \rightarrow \langle \epsilon, \tree', \node', ctx', i' \rangle \+
ctx \neq ctx'
}
{\langle op, \tree, \node, ctx, i \rangle \xrightarrow{} \langle l_{false}, \tree, \node, ctx, i \rangle}
\\
\end{array}
\]
\end{minipage}
\end{tabular}

\caption{\tlang{} language small-step semantics. Each rule is of the type: $\tlang{} \times States \rightarrow \tlang{} \times States$.
}
\figlabel{tlangsemantics}

\end{figure}

Move instructions are described by the \textsc{[Move]} and \textsc{[Move-Fail]} rules in \figref{tlangsemantics} and use the function $mv$:
\[mv \colon \scode{Move} \times AST \times X \times \mathbb{I}^* \rightarrow X \times \mathbb{I}^*\]
defined as follows:

\begin{itemize}
\item $mv(\scode{\color{red!50!black}Up}, \tree, \node, i) = \node' \times i$ where $\node'$ is the parent node of $\node$ in $\tree$ or $\bottom$ if $\node$ has no parent node in $\tree$. Note that the \textsc{[Move]} rule updates the node at the current position to be the parent.

\item $mv(\scode{\color{red!50!black}Left}, \tree, \node, i) = \node' \times i$ where $\node'$ is the left sibling of $\node$ in $\tree$ or $\bottom$ if $\node$ has no left sibling. Similarly, $mv(\scode{\color{red!50!black}Right}, \tree, \node)$ produces the right sibling or $\bottom$ if $\node$ has no right sibling.

\item $mv(\scode{\color{red!50!black}DownFirst}, \tree, \node, i) = \node' \times i$ where $\node'$ is the first child of $\node$ in $\tree$ or $\bottom$ if $\node$ has no children. Similarly, $mv(\scode{\color{red!50!black}DownLast}, \tree, \node, i)$ produces the last child of $\node$ or $\bottom$ if $\node$ has no children.

\item $mv(\scode{\color{red!50!black}GoToGlobal}, \tree, \node, i) = \node' \times i$ where $\node'$ is a node corresponding to the \scode{global} JavaScript object in the $\tree$. For this and other \scode{GoTo} operations the value of $\node'$ is independent of the starting node $\node$.

\item $mv(\scode{\color{red!50!black}GoToThis}, \tree, \node, i) = \node' \times i$ where $\node'$ is a node corresponding to the object to which \scode{this} keyword points-to in the top-level scope. In a web browser this would be \scode{window} object while in \scode{Node.js} application it is \scode{module.exports}.

\item $mv(\scode{\color{red!50!black}GoToUndefined}, \tree, \node, i) = \node' \times i$ where $\node'$ is a node corresponding to the \scode{undefined} JavaScript object in the $\tree$. Similarly, for $mv(\scode{\color{red!50!black}GoToNull}, \tree, \node, i) = \node' \times i$ the $\node'$ is the \scode{null} value.

\item $mv(\scode{\color{red!50!black}GoToCaller}, \tree, \node, i \cdot i') = \node' \times i'$ where $\node'$ is the node corresponding to call site of the top method $i$ from call trace and $i'$ is the call trace with the method $i$ removed. If the call trace is empty then $\node' = \bottom$.

\item $mv(\scode{\color{red!50!black}UpUntilFunc}, \tree, \node, i) = \node' \times i$ navigates recursively to the first $\node'$ using the $\scode{\color{red!50!black}Up}$ operation such that the parent of $\node'$ is a function declaration or root of the tree is reached.

\item $mv(\scode{\color{red!50!black}Top}, \tree, \node, i) = \top \times i$ denotes that the analysis approximates the result to the $\top$ element in the lattice.

\end{itemize}

\subsection{\tlang{}: Small-step Semantics of \tlang{} language.}

Recall from \appref{DSLlanguage} that \tlang{} programs operate on a state $\sigma$ defined as follows: $\sigma = \langle \tree, \node, ctx, i\rangle \in States$ where the domain $States$ is defined as $States = AST \times X \times Context \times \mathbb{I}^*$.
Initially, execution starts with the empty context $[] \in Context$ and for a program $p \in \tlang{}$, a tree $\tree \in AST$, and a position $\node \in X$ in the tree, we say that program $p$ computes a position $\node' \in X$, denoted as $p(\tree, \node, i) = \node'$, iff there exists a~sequence of transitions from $\langle p, \tree, \node, [], i \rangle$~to~$\langle \epsilon, \tree, \node', [], i \rangle$. The small-step semantics of executing a \tlang{} program are shown in \figref{tlangsemantics}.

\section{Allocation Site Analysis}\seclabel{allocsite}
In this section we describe the instantiation of our approach to the task of learning allocation site analysis. The goal of allocation site analysis is to answer queries of the type $q: L \rightarrow \lbrace true, false \rbrace$, where $L$ is a set of program locations. That is, for each program location the analysis returns a boolean value denoting whether the location is an allocation site or not.

\paragraph{Our Goal}
Our goal is to learn an inference rules from data in the following shape:

\[
\footnotesize
\inferrule*[Right=\footnotesize \textsc{$~$[Alloc]}]
{f(l) = true}
{\scode{AllocSite(}l\scode{)}}
\]

\begin{figure}[t]
\center
\begin{tikzpicture}

\node at (0, 0) {
\footnotesize
\begin{minipage}{.65\textwidth}
\centering

\begin{lstlisting}[language=JavaScript, tabsize=2, basicstyle=\ttfamily\small, keepspaces=true]
var obj = _{a: 7}_;
var arr = _[1, 2, 3, 4]_;
if (obj.a == _arr.slice(0,2)_) { ... }
var n = _new Number(7)_;
var obj2 = new Object(obj);
try {  ...  } catch (_err_) {  ...  }
\end{lstlisting}
\end{minipage}
};

\node[text width=6cm, align=center] at (6, 0) {\underline{Allocation Sites} \\(new object allocated)};


\end{tikzpicture}
\vspace{-2.5em}
\caption{Illustration of program locations (underlined) for which the allocation site analysis should report that a new object is allocated.}

\figlabel{alloc_example}
\end{figure}
\paragraph{Example}
We illustrate the expected output and some of the complexities of allocation site analysis on a small example shown in \figref{alloc_example}.
The goal of the analysis is to determine all the program locations at which new object is allocated. 
In JavaScript there are various ways how an object can be allocated, some of which are shown in \figref{alloc_example}. These include creating new object without calling a~constructor explicitly (for example by creating new array or object expression inline), creating new object by calling a~constructor explicitly using \scode{new}, creating a new object by calling a~method or new objects created by throwing an exception.
In addition, some of the cases might further depend on actual values passed as arguments. For example, calling a \scode{new Object(obj)} constructor with \scode{obj} as an argument does not create a new object but returns the \scode{obj} passed as argument instead.

\subsection{Instantiating our Learning Approach}
We now define the necessary components required to instantiate the learning approach described in our work.

\tikzset{
    >=stealth',
    pil/.style={
           thin,
           shorten <=2pt,
           shorten >=2pt,}
}

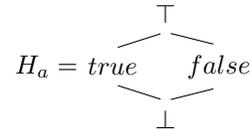
\begin{wrapfigure}[7]{r}{0.35\textwidth}
\centering
\vspace{-2.75em}
\begin{tikzpicture}
\node[] (bottom) {$\bot$};
\node[] (top) at ([yshift=40pt]bottom) {$\top$};

\node[] (h2) at ([xshift=-20pt, yshift=20pt]bottom) {$true$};
\node[] (h4) at ([xshift=20pt, yshift=20pt]bottom) {$false$};

\node[] at ([xshift=-25]h2) {$H_a =$};

\draw[pil] (bottom.north) -- (h2.south);
\draw[pil] (bottom.north) -- (h4.south);

\draw[pil] (h2.north) -- (top.south);
\draw[pil] (h4.north) -- (top.south);

\end{tikzpicture}
\caption{Lattice used for allocation site analysis.}
\figlabel{alloc_lattice}
\end{wrapfigure}

\paragraph{Abstract Lattice}
\figref{alloc_lattice} shows the lattice $(\mathcal{H}_{a}, \sqsubseteq)$ used to represent the abstract domain for allocation site analysis.
The abstraction function $\alpha \colon L \rightarrow H_a$ maps the concrete program locations to elements $true$ and $false$ which denote whether the program location is an allocation site or not.

\paragraph{Concrete and abstract program semantics}

The concrete properties we are tracking and their abstract counterpart as described in \secref{analysisCorrectness} are instantiated by setting $\mathcal{C} := \lbrace true,  false \rbrace$, $\mathcal{A} := H_a$ and $\mathbb{N} := L$, where $L$ is a set of all program locations (nodes in an AST).
That is, all concrete program behaviors are captured by a function $\ConcSem{p} \colon \langle L \rangle \rightarrow \lbrace true,  false \rbrace$ that for each program location $L$ computes whether it is an allocation site.
The abstract semantics are similar except that we instantiate the abstract domain to be the lattice $(\mathcal{H}_{a}, \sqsubseteq)$.
We discuss how we obtain the concrete behaviors $\ConcSem{p}_{ti}$ after running the program on a set of test inputs $ti$ in \appref{obtainingTrace}.

\paragraph{Program modifications}

We use the same set of program modification as used to learn points-to analysis (described in \appref{pointsto}).

\begin{figure}[t]
{

\begin{align*}
m \in \scode{Move}_{\scode{core}} ::=&~ \scode{{\color{red!50!black}Up} | {\color{red!50!black}Left} | {\color{red!50!black}Right} | {\color{red!50!black}DownFirst} | {\color{red!50!black}DownLast} | {\color{red!50!black}Top}}\\[1pt]
m \in \scode{Move}_{\scode{alloc~~}} ::=&~ \scode{{\color{red!50!black}PrevNodeValue} | {\color{red!50!black}PrevNodeType} }\\[1pt]
m \in \scode{Move} ::=&~ \scode{Move}_{\scode{core}}~\cup~\scode{Move}_{\scode{alloc}} \\[1pt]
w \in \scode{Write} ::=&~ \scode{{\color{blue!50!black}WriteValue}\,|\,{\color{blue!50!black}WritePos}\,|\,{\color{blue!50!black}WriteType}\,|\,{\color{blue!50!black}HasPrevNodeValue} } \\[1pt]
&~ \scode{{\color{blue!50!black}NewAlloc}\,|\,{\color{blue!50!black}NoAlloc}} 
\end{align*}

%
}
\vspace{-1.6em}
\caption{Language \alang{} for expressing the result of allocation site query by means of navigating over an abstract syntax tree.}
\figlabel{alloc_language}
\end{figure}

\paragraph{Language for allocation site analysis}
The DSL language \alang{} used to instantiate the learning of allocation site analysis is very similar to the \tlang{} used for points-to analysis.
The syntax of the language is shown in \figref{alloc_language} and is based on the same idea of navigating over the abstract syntax tree of a given program.
It contains two additional instructions \scode{\color{blue!50!black}NewAlloc} and \scode{\color{blue!50!black}NoAlloc} used to denote whether a given location is an allocation site. These two instructions are used in the leafs of the learned analysis.
Additionally, it defines two general instructions used to navigate over the AST -- \scode{\color{red!50!black}PrevNodeValue} and \scode{\color{red!50!black}PrevNodeType}. The formal semantics of these instructions are following:

\begin{itemize}

\item $mv(\scode{\color{red!50!black}PrevNodeValue}, \tree, \node) = \node'$ where $\node'$ is 
the position of the most recent AST node with the same value as the current node, \ie, the maximal $\node'$ such that $\node' < \node$ and $wr(\scode{\color{blue!50!black}WriteValue}, \tree, \node) = wr(\scode{\color{blue!50!black}WriteValue}, \tree, \node')$. Further, to enable modular learning we require that both nodes $\node$ and $\node'$ are defined within the same function or in the top level scope of the program. If no such value $\node'$ exists in the $\tree$ then a $\bottom$ is returned.

\item $mv(\scode{\color{red!50!black}PrevNodeType}, \tree, \node) = \node'$ has the same semantics as \scode{\color{red!50!black}PrevNodeValue} except that we require that the types at given nodes are the same, \ie, $wr(\scode{\color{blue!50!black}WriteType}, \tree, \node) = wr(\scode{\color{blue!50!black}WriteType}, \tree, \node')$.

\end{itemize}

Finally, we note that the formal semantics of the \alang{} are the same as presented for \tlang{} except that for \alang{} we do not include the information about current call trace. That is, the program state $\sigma$ is defined as follows: $\sigma = \langle \tree, \node, ctx\rangle \in States$ where the domain $States = AST \times X \times Context$.
\section{Learned Program Analyses}\seclabel{examples}

\subsection{Points-to Analysis}

To illustrate the complexity of the learned program analysis and the fact that it is easy for it to be interpreted by a human expert, we show the learned analysis for the API \scode{Array.prototype.filter} in \figref{filter_analysis}.
By inspecting the programs in the branches we can see that the analysis learns three different locations in the program to which the \scode{this} object can point-to: the \scode{global} object, a newly allocated object, or the second argument provided to the \scode{filter} function. The analysis also learns the conditions determining which location to select. For example, \scode{this} points to a new allocation site only if the second argument is a primitive value, in which case it is boxed by the interpreter. Similarly, \scode{this} points-to the second argument (if one is provided), except for cases where the second argument is \scode{null} or \scode{undefined}.

For better readability we replaced the sequence of instructions in \tlang{} used as branch conditions and branch targets with their informal descriptions. For example, the learned sequence that denotes the second argument of the calling method is $\color{red!50!black} \scode{GoToCaller DownFirst Right Right}$.
It is important to note that that we were not required to manually provide any such sequences in the language but that the learning algorithm discovered such relevant sequences automatically.

\begin{figure}[t]
\center

\begin{tabular}{l}
$\footnotesize \texttt{Array.prototype.filter} ::= $\\
~~$\footnotesize \bcode{if}~~\texttt{caller has one argument}~~\bcode{then}$\\
~~~~$\footnotesize \texttt{points-to global object}$\\
~~$\footnotesize \bcode{else if}~~\texttt{2nd argument is Identifier}~~\bcode{then}$\\
~~~~$\footnotesize \bcode{if}~~\texttt{2nd argument is undefined}~~\bcode{then}$\\
~~~~~~$\footnotesize \texttt{points-to global object}$\\
~~~~$\footnotesize \bcode{else}$\\
~~~~~~$ \footnotesize\texttt{points-to 2nd argument}$\\
~~$\footnotesize \bcode{else if}~~\texttt{2nd argument is This}~~\bcode{then}$\\
~~~~$\footnotesize \texttt{points-to 2nd argument}$\\
~~$\footnotesize \bcode{else if}~~\texttt{2nd argument is null}~~\bcode{then}$\\
~~~~$\footnotesize \texttt{points-to global object}$\\
~~$\footnotesize \bcode{else}~~\texttt{//2nd argument is a primitive value}$\\
~~~~$\footnotesize \texttt{points-to new allocation site}$\\
\end{tabular}
\caption{Learned analysis for JavaScript API \scode{Array.prototype.filter}.}
\figlabel{filter_analysis}
\end{figure}

\subsection{Allocation Site Analysis}
Here we provide details and the learned program for allocation site analysis.
We start by describing a subset of the learned program that corresponds to handling of statements that allocate objects using \scode{NewExpression}. Then we describe the full analysis.

\paragraph{Program learned for object allocation using \scode{NewExpression}}
As illustrated in \figref{alloc_example}, calling \scode{new} in a JavaScript program does not necessarily lead to allocation of a new object. The exception are the semantics of the built-in \scode{Object} class that are defined as follows\footnote{\url{https://developer.mozilla.org/en-US/docs/Web/JavaScript/Reference/Global_Objects/Object}}:

\begin{center}
\textit{
``The \scode{Object} constructor creates an object wrapper for the given value. If the value is \scode{null} or \scode{undefined}, it will create and return an empty object, otherwise, it will return an object of a Type that corresponds to the given value. If the value is an object already, it will return the value.``
\rightline{{\rm --- \scode{Object constructor}}}}
\end{center}

By inspecting the learned program shown in \figref{new_alloc_analysis} we can see that it learns the above semantics by checking the type of the argument passed to the constructor. 
If the argument is one of the primitive types or a \scode{null} value then it will be always wrapped in an new object (marked by returning \scode{NewAlloc} as the leaf program). The program also learns that in case the constructor has no arguments it always allocates a new object.
In case the argument was used before then the analysis chooses to conservatively approximate the result.

\begin{figure}[t]
\center

\begin{tabular}{l}
~~$\footnotesize \bcode{if}~~\texttt{WriteType == NewAllocation}~~\bcode{then}$\\
~~~~$\footnotesize \bcode{if}~~\texttt{constructor for given object was used before}~~\bcode{then}$\\
~~~~~~$\footnotesize \texttt{NewAlloc}$\\
~~~~$\footnotesize \bcode{else if}~~\texttt{last argument is LiteralNumber}~~\bcode{then}$\\
~~~~~~$\footnotesize \texttt{NewAlloc}$\\
~~~~$\footnotesize \bcode{else if}~~\texttt{last argument is LiteralString}~~\bcode{then}$\\
~~~~~~$ \footnotesize\texttt{NewAlloc}$\\
~~~~$\footnotesize \bcode{else if}~~\texttt{constructor with no arguments}~~\bcode{then}$\\
~~~~~~$ \footnotesize\texttt{NewAlloc}$\\
~~~~$\footnotesize \bcode{else if}~~\texttt{last argument is LiteralBoolean}~~\bcode{then}$\\
~~~~~~$ \footnotesize\texttt{NewAlloc}$\\
~~~~$\footnotesize \bcode{else if}~~\texttt{last argument is UnaryExpression}~~\bcode{then}$\\
~~~~~~$ \footnotesize\texttt{NewAlloc}$\\
~~~~$\footnotesize \bcode{else if}~~\texttt{last argument is ArrayExpression}~~\bcode{then}$\\
~~~~~~$ \footnotesize\texttt{NewAlloc}$\\
~~~~$\footnotesize \bcode{else if}~~\texttt{last argument is null}~~\bcode{then}$\\
~~~~~~$ \footnotesize\texttt{NewAlloc}$\\
~~~~$\footnotesize \bcode{else}$\\
~~~~~~$\bcode{if}~~\texttt{last argument has been used before}~~\bcode{then}$\\
~~~~~~~~$\footnotesize\texttt{Top}$\\
~~~~~~$\footnotesize \bcode{else}$\\
~~~~~~~~$\footnotesize\texttt{Top}$\\
\end{tabular}
\caption{Learned analysis for object allocation by invoking the constructor explicitly.}
\figlabel{new_alloc_analysis}
\end{figure}

\paragraph{Full learned analysis}

The summary of the full allocation site analysis learned for JavaScript is shown in \figref{alloc_analysis}. We can see that the analysis iteratively refines the dataset by conditioning on various types of predictions.

First, the analysis checks whether the given value was used before in the program. This case typically applies to global objects for which their first reference in the program is considered an allocation site. Therefore, if the object was seen before (\ie, \scode{HasPrevNodeValue} returns $true$) it is very likely that it is not an allocation site. However this might not be always the case and therefore the learning algorithm chooses to approximate this branch (as it cannot find a~further refinement). A counter-example for using \scode{NoAlloc} program inside this branch is for example a program that uses \scode{new Map()} and then \scode{Map.prototype}. Here, even though the \scode{Map} is used before in the program, it is considered a new allocation site at the time the \scode{prototype} field is accessed. This is because the instrumentation does not track the read of object \scode{Map} when calling a constructor.

Subsequently, the analysis identifies that calling some methods might return a new object and learns to model such cases. Here an interesting first branch that is learned is to check whether the call is used in an \scode{ExpressionStatement}, \ie, as a single statement. In this case the return value is not used in the program and therefore is unlikely to be an allocation site. However, similar to the previous case, this is not guaranteed and therefore the algorithm learns to approximate this branch.

Next, an analysis or accessing elements in an array is learned. Note that this analysis is quite complex as the elements in the array might alias with other variables which makes it difficult for the analysis to precisely determine a simple model for this case.

Further, a simple model is learned for the standard allocation site locations such as the arguments and implicit constructors. Finally, the analysis also learns that the left hand side of an assignment cannot be an allocation site.

\begin{figure}[h]
\center

\begin{tabular}{l}
~~$\footnotesize \bcode{if}~~\texttt{HasPrevNodeValue}~~\bcode{then}$\\
~~~~$\footnotesize \texttt{Top}$\\
~~$\footnotesize \bcode{else if}~~\texttt{WriteType == CallExpression}~~\bcode{then}$\\
~~~~$\footnotesize \bcode{if}~~\texttt{Up WriteType == ExpressionStatement}~~\bcode{then}~~\texttt{//return value not assigned}$\\
~~~~~~$\footnotesize \texttt{Top}$\\
~~~~$\footnotesize \bcode{else}$\\
~~~~~~$ \footnotesize\texttt{...}$\\
~~$\footnotesize \bcode{else if}~~\texttt{WriteType == ArrayAccess}~~\bcode{then}$\\
~~~~$ \footnotesize\texttt{...}$\\
~~$\footnotesize \bcode{else if}~~\texttt{Up WriteType == CatchClause|FunctionExpression}~~\bcode{then}$\\
~~~~$\footnotesize \texttt{NewAlloc}~~\texttt{//arguments}$\\
~~$\footnotesize \bcode{else if}~~\texttt{WriteType == ObjectExpression|ArrayExpression|LiteralRegExp}~~\bcode{then}$\\
~~~~$\footnotesize \texttt{NewAlloc}~~\texttt{//implicit constructors}$\\
~~$\footnotesize \bcode{else if}~~\texttt{WriteType == NewExpression}$\\
~~~~$\footnotesize \texttt{...}$\\
~~$\footnotesize \bcode{else if}~~\texttt{Up WriteType == AssigmentExpression}$\\
~~~~$\footnotesize \bcode{if}~~\texttt{left hand side of the assigment}~~\bcode{then}$\\
~~~~~~$\footnotesize \texttt{NoAlloc}$\\
~~~~$\footnotesize \bcode{else}$\\
~~~~~~$\footnotesize \texttt{...}$\\
~~$\footnotesize \bcode{else}$\\
~~~~$\footnotesize \texttt{...}$\\
\end{tabular}
\caption{Summary of learned allocation site analysis for JavaScript.}
\figlabel{alloc_analysis}
\end{figure}

\section{Implementation}
In this section we describe the implementation details of our approach.

\subsection{Obtaining Programs's Concrete Behaviors $\ConcSem{p}_{ti}$}\seclabel{obtainingTrace}

We extract the relevant concrete behaviours of the program $p$ by instrumenting the source code (not the interpreter) such that when executed, $p$ produces a~trace $\pi$ consisting of all object reads, method enters, method exits and call sites. Additionally, at each method entry, we record the reads of all the parameters and the value of \scode{this}. Further, every element in the trace contains a mapping to the location in the program (in our case to the corresponding node in the AST) and object reads record the unique identifier of the object being accessed. 

\paragraph{Training dataset for points-to analysis}
Given such a trace $\pi$, we create a~dataset $\dataset_{pt}$ used for points-to analysis by generating one input/output example for each position in the trace $\pi$ at which the \scode{this} variable was read. Further, we select only the first read of \scode{this} in each scope as all such references point to the same object. An input/output example is a pair $\langle AST \times X \times \mathbb{I}^*, O \rangle$, where $t \in AST$ is an abstract syntax tree corresponding to the input program, $n \in X$ is a position in the tree where a given read was performed, $i \in \mathbb{I}^*$ is a call trace and $o \in O$ is the identifier of the concrete object seen during execution.



\paragraph{Training dataset for allocation site analysis}
Given such a trace $\pi$, we create a~dataset $\dataset_{alloc}$ used for allocation site analysis by generating one input/output example for each position in the trace $\pi$ as follows:
\begin{enumerate}
\item select all positions in the trace $\pi$ where an object was read.
\item for each position select only the first read in the trace, \ie, first loop iteration or first method invocation.
\item filter reads of \scode{this} object and field access.

\end{enumerate}

An input/output example is a pair $\langle AST \times X, O \rangle$, where $t \in AST$ is an abstract syntax tree corresponding to the input program, $n \in X$ is a position in the tree where a given read was performed.
From a trace $\pi$ we determine the correct label $O \in \lbrace true, false \rbrace$ by assigning label $true$ for all positions in $\pi$ for which the corresponding identifier of the object being accessed was not seen previously within the same method call (or global scope) and $false$ otherwise. That is, intuitively we say that program location is an allocation site if the object being read was not seen before. We consider method call boundaries to make the analysis modular an independent of the current program call trace.

\subsection{Checking Analysis Correctness}\seclabel{implCorrectness}

\paragraph{Points-to analysis}
For a program analysis $\analysis$ and a dataset $\dataset$, we are interested in checking whether the analysis results computed for program $p$ are correct with respect to the concrete values seen during the execution of $p$.
Recall (from \appref{DSLlanguage}) that executing the analysis $\analysis \in \tlang{}$ on an input example $\langle t , \node, i, o \rangle$ (as defined above) produces a position $\node' = \analysis(t, \node, i)$ in the program or the element~$\top$. If the analysis returns $\top$ then it is trivially correct, otherwise we distinguish between two cases.
If $\node' = \node$, we say that the analysis is correct if the value $o$ has not been seen in the trace $\pi$ before position $\node$. This is true when position $\node$ is a new allocation site.
If $\node' \neq \node$, we say that the analysis is correct if the value $o$ has been seen previously in the trace at position $\node'$.

\paragraph{Allocation site analysis}
Checking the correctness of the allocation site analysis is trivial as executing the analysis $\analysis \in \alang{}$ on an input example $\langle t , \node, o \rangle$ produces one of the labels \scode{NewAlloc}, \scode{NoAlloc} or \scode{Top} which can be directly compared to the expected output $o \in \lbrace true, false \rbrace$.

\subsection{Synthesising \tlang{} and \alang{} Programs}

We instantiate the learning described in \secref{learning} using the following two program generators \textit{genAction} and \textit{genBranch} for the \tlang{} and \alang{} languages. For \tlang{} we instantiate \textit{genAction} using an enumerative search that considers all programs up to size~$5$. For \alang{} the \textit{genAction} simply tries two possible programs  \scode{NewAlloc} and \scode{NoAlloc}.

We instantiate the \textit{genBranch} using the same procedure for both \tlang{} and \alang{} languages.
In particular, we use enumerate search that considers as conditions all programs up to size $6$ (with up to $5$ move and $1$ write instruction). To determine the concrete value used as a right-hand side of the condition, we collect the top $10$ most common values observed when executing the condition and pick one that maximizes the information gain metric as defined in \secref{id3}.
For the dataset sizes used in our work such simple generators proved to be effective in practice. To scale for larger datasets one could use the idea of representative sampling \cite{NoiseLearning} that was shown to work well for the domain of programs.

\subsection{Regularization}
The purpose of regularization is to select simpler programs from language $\lang$. In particular, we use the following regularized cost function $cost_{reg}(\dataset, \analysis) = cost(\dataset, \analysis) + \lambda \cdot \Omega(\analysis)$, where $\lambda$ is a~regularization constant empirically set to $0.01$ and $\Omega(\analysis)$ is a~regularization that penalizes more complex programs. We instantiate $\Omega(\analysis)$ to return number of instructions in $\analysis$.
Additionally, for programs that use \scode{WritePos} and \scode{WriteValue} we multiply the regularization $\Omega(\analysis)$ by a factor of two as these values are less stable under program modification.
We note that using such regularized cost function directly in the learning is a useful extension of our approach that allows controlling the amount of approximation (by setting the value of~$\lambda$).

\subsection{JavaScript Restrictions}
Finally, we remove from the training data programs that use the \scode{eval} function, dynamic function binding using \scode{Function.prototype.bind} and \scode{Function} object constructor.
These are language features that require combination of analyses to handle precisely and are therefore typically ignored by static analyzers~\cite{Livshits:2015}.
We also filter accesses to \scode{arguments} object for the allocation site analysis. This is a limitation of our instrumentation that instruments reads and methods calls by means of wrapper functions that affect the binding of \scode{arguments} object.

\end{document}